\documentclass{aa}  

\usepackage{graphicx}
\usepackage{txfonts}
\usepackage{mathabx}
\usepackage{tabularx}
\usepackage{booktabs}
\usepackage[justification=centering]{subcaption}
\usepackage{multirow}
\usepackage[colorlinks=true, allcolors=blue]{hyperref}
\usepackage[noabbrev, capitalise]{cleveref}
\usepackage{natbib}
\bibpunct{(}{)}{;}{a}{}{,} 

\makeatletter
\renewcommand*\aa@pageof{, page \thepage{} of \pageref*{LastPage}}
\makeatother

\newcolumntype{L}{>{\raggedright\arraybackslash}X}
\newcolumntype{R}{>{\raggedleft\arraybackslash}X}

\begin{document}

   \title{Exoplanets across galactic stellar populations with PLATO}
   \subtitle{Estimating exoplanet yields around FGK stars for the thin disk, thick disk, and stellar halo}

   \author{C. Boettner
          \inst{1}
          \fnmsep
          \inst{2}
          \and
          A. Viswanathan 
          \inst{1}
          \and
          P. Dayal
          \inst{1}
          }

   \institute{Kapteyn Astronomical Institute, University of Groningen,
              Landleven 12 (Kapteynborg, 5419) 9747 AD Groningen\\
              \email{boettner@astro.rug.nl}
              \and
              GELIFES Institute, University of Groningen,
              Nijenborgh 7 9747 AG Groningen
             }

   \date{Received xxx; accepted xxx}
 
  \abstract
   {The vast majority of exoplanet discoveries to date have occurred around stars in the solar neighbourhood, with chemical compositions comparable to that of the Sun. However, models suggest that planetary systems in different Galactic environments, with varying dynamical histories and chemical abundances, may exhibit distinct characteristics, which can help  improve our understanding of planet formation processes.}
   {This study aims to assess the potential of the upcoming PLATO mission to investigate exoplanet populations around stars in diverse Galactic environments, specifically focusing on the Milky Way thin disk, thick disk, and stellar halo. We aim to quantify PLATO's ability to detect planets in each environment and determine how these observations could constrain planet formation models.}
   {Beginning with the all-sky PLATO Input Catalogue,
    we kinematically classified the 2.4 million FGK stars into their respective Galactic components. For the sub-sample of stars in the long-observation LOPS2 and LOPN1 PLATO fields, we estimated planet occurrence rates using the New Generation Planet Population Synthesis (NGPPS) dataset. Combining these estimates with a PLATO detection efficiency model, we predicted the expected planet yields for each Galactic environment during a nominal 2+2 year mission.}
   {Based on our analysis, PLATO is likely to detect at least 400 exoplanets around the $\alpha$-enriched thick disk stars. The majority of those planets are expected to be super-Earths and sub-Neptunes with radii between 2 and 10 $R_\Earth$ and orbital periods between 2 and 50 days, which is ideal for studying the link between the radius valley and stellar chemistry. For the metal-poor halo, PLATO is likely to detect between 1 and 80 planets with periods between 10 and 50 days, depending on the potential existence of a metallicity threshold for planet formation. The PLATO fields contain more than 3,400 potential target stars with [Fe/H] < -0.6, which will help  improve our understanding of planets around metal-poor stars. We identified a specific target list of 47 (kinematically classified) halo stars in the high-priority, high-signal-to-noise PLATO P1 sample, offering prime opportunities in the search for planets in metal-poor environments.}
   {PLATO's unique capabilities and large field of view position it as a valuable 
    tool for studying planet formation across the diverse Galactic environments of 
    the Milky Way. By probing exoplanet populations around stars with a varying 
    chemical composition, PLATO will provide helpful insights into the connection 
    between stellar chemistry and planet formation.}

   \maketitle
%
\section{Introduction}
\label{sec:introduction}
The properties of planetary systems are known to be linked to those of their host stars. This relationship is particularly evident in the well-established correlation between the stellar metallicity and prevalence of giant planets. Close-in giant planets are more likely to be found around metal-rich stars \citep{Gonzalez1997, Fischer2005, Johnson2010, Buchhave2012, Thorngren2016a}, and giants in general have lower eccentricities when orbiting metal-poor stars \citep{Dawson2013, Buchhave2018}. Further, the characteristic radius valley -- a decrease in the occurrence rates of planets between super-Earths and sub-Neptunes \citep{Fulton2017} -- also appears to evolve over time. While warm super-Earths with periods between 10–100 days have a nearly constant occurrence rate of -0.4 < [Fe/H] < 0.4, warm sub-Neptune occurrence rates double over this range \citep{Petigura2018a}. Orbital periods of small planets around metal-poor stars tend to be longer \citep{Beauge2013, Wilson2018}, whereas close-in ($P < 10$ days) small ($< 2~M_\Earth$) planets are more likely to be found around metal-rich stars \citep{Mulders2016, Lu2020}. In addition to metallicity, planet properties also correlate with stellar age, effective temperature, and mass. According to \citet{Howard2012} and \citet{Mulders2015}, the number of close-in planets, ranging from Earth- to Neptune-sized, decreases as stellar mass and temperature increase. \citet{Yang2020} and  \citet{He2021} report that the fraction of stars hosting planets changes from roughly 0.3 for F-type stars to around 0.8 for K-type stars. 

Across the Milky Way, stellar populations exhibit significant variations in age and chemical composition \citep{Gilmore1983, Prochaska2000, Reddy2006, Schuster2006, Bensby2014, Masseron2015}. This suggests that the local Galactic environment may significantly impact planetary populations \citep{Nielsen2023, Boettner2024}. Within the solar neighbourhood, distinct stellar populations can be classified into three categories: the thin disk, which is generally composed of young stars with roughly solar-like metallicities; the thick disk, which hosts an older, more metal-poor but noticeably $\alpha$-enriched population \citep{Wallerstein1962}; and the halo, which hosts some of the oldest and most metal-poor stars, with metallicities ranging from -7 < Fe/H < -0.5 \citep{Beers2005, Frebel2015, Nordlander2017, Starkenburg2018}. The majority of exoplanet studies have so far been focused on planets orbiting nearby thin disk stars \citep{Bashi2019}, primarily with stellar [Fe/H] > -0.5 and solar-like $\alpha$ abundances. These stars are more common in the solar neighbourhood, and on average are brighter than their older counterparts, naturally promoting their planetary companions for transit and radial velocity detections. Because of this, the effects of the Galactic environment on planet demographics has so far been largely unexplored.

More recently, however, the impact of stellar abundance variations on planet formation has attracted growing interest. Theoretical studies by \citet{Bitsch2020} and \citet{Cabral2023} have analysed the effects of chemical abundances from the GALAH and APOGEE spectroscopic surveys on the chemical composition of planetary building blocks (PBBs). \citet{Bitsch2020} find that while PBB composition inside the water ice line (T > 150K) does not change significantly with increasing [Fe/H], outside the ice line, water ice content drastically decreases from nearly 50\% at [Fe/H] = -0.4 to 6\% at [Fe/H] = 0.4. This decrease is caused by an increasing C/O ratio that binds oxygen in gaseous CO and CO$_{2}$. They predict that planets that formed outside the ice line will have a strong dependence of water ice content on host star metallicity, with more metal-poor stars hosting more water-rich planets.

Some studies suggest that a high $\alpha$ abundance directly affects the chemical composition of the PBBs, even at a fixed [Fe/H] value. Larger $\alpha$ contents may lead to more icy PBBs, which aids planet formation and can partially compensate for low [Fe/H] metallicity \citep{Bashi2019, Bashi2020, Morbidelli2015, Gundlach2015}. \citet{Cabral2023} have found the water content of PBBs to be decreasing with increasing [Fe/H], and that the bimodal distribution of stars in the [Fe/H] - [$\alpha$/Fe] plane \citep{Wallerstein1962} leads to a bimodal distribution in PBB compositions, where most chemical species have a $\alpha$-content dependence. Similar results were found by \citet{Santos2017} and \citet{Cabral2019}, suggesting that PBB chemical composition might strongly depend on the $\alpha$ abundance and thus the Galactic origin. However, the exact effect of $\alpha$ elements is not entirely clear. Based on the APOGEE sample, \citet{Cabral2023} found that thin disk stars host more water-rich PBBs, while their GALAH sample showed the opposite. Furthermore, whether the water abundance of PBBs is a good proxy for planetary water contents is still debated. Observationally, \citet{Bashi2022} found that close-in super-Earths are more likely to be hosted around thin disk stars rather than thick disk stars, but their sample was limited to only eight planets that were found around thick disk hosts. \citet{Adibekyan2012} have found that planet-hosting stars have a higher abundance of $\alpha$ elements, and \citet{Chen2022} further suggested that the ratio of super-Earths to sub-Neptunes decreases significantly with increasing [$\alpha$/Fe] in stars, meaning sub-Neptunes are more common around thick disk stars.
\citet{Zink2023} analysed the occurrence rate of planets around FGK dwarf stars observed by the K2 mission, finding that stars with large galactic oscillation amplitudes host fewer super-Earths and sub-Neptunes. They demonstrate that the metallicity gradient with galactic scale height alone cannot explain this observed trend, and suggest that the increased relative alpha-element abundance of these high amplitude stars may influence planet formation. It has also been suggested that while giant planets around sub-solar metallicity stars are rare, they are more common around thick disk stars \citep{Haywood2008, Haywood2009}.

The detection of a large population of close-in giant planets orbiting high-metallicity stars has drawn considerable research attention to such high-metallicity systems since the early days of exoplanet discoveries \citep{Santos2004, Fischer2005, Johnson2010a, Mortier2013, Petigura2018a}. As a result, studies of low-metallicity systems have largely played a secondary role. However, this focus has begun to shift in recent years, with increasing attention being paid to planets around low-metallicity stars. A long-standing idea is the existence of a minimum metallicity requirement for planet formation. \citet{Johnson2012a} argued for a critical metallicity of [Fe/H]$_\mathrm{crit}\sim -1.5 + \log(r/1 \mathrm{AU})$. Observationally, similar limits have been proposed around [Fe/H] = -0.5 \citep{Mortier2012, Mordasini2012} and -0.7 \citep{Boley2021}. Specifically for hot Jupiters ($P$ < 10 days, $R$ between 0.8–2 $R_\mathrm{Jupiter}$), \citet{Boley2021} found an upper limit on the occurrence rate of 0.18\% for stars with -2 < [Fe/H] < -0.6 based on a study of 11,125 halo stars from the TESS survey. The lowest known metallicity of a hot Jupiter host is [Fe/H] = -0.6 \citep{Hellier2014}. A suggested reason for this is the shortened protoplanetary disk lifetime in a metal-poor region (\citealt{Kornet2005, Ercolano2010, Yasui2010}, see however \citealt{Winter2024}). \citet{Andama2024} studied planetesimal formation in low-metallicity environments using the 1D \texttt{chemocomp} code. They found that the amount of planetesimals (in terms of mass) decreases with decreasing disk and host star metallicity, especially for low-viscosity disks. The total planetesimal mass drops by multiple orders of magnitude when going from [Fe/H] = 0 to [Fe/H] = -0.4. The drop is weaker for slightly higher viscosity, more turbulent disks, but it is still about one order of magnitude, with a lower limit of [Fe/H] = -0.6. Low metallicity environments therefore present an ideal environment to study the early stages of planet formation.

The PLAnetary Transits and Oscillations of stars (PLATO) mission is a planned European Space Agency space telescope set to launch in 2026, with the primary mission target being the search for terrestrial exoplanets in the habitable zones of Sun-like stars using the transit method. PLATO's unique design, featuring 24 `normal` cameras and two `fast` cameras with a higher cadence \citep{Rauer2014a, Rauer2016}, covers a field of view (FoV) of $49^\degree \times 49^\degree$ per pointing, which is about 20 times larger than \textit{Kepler} \citep{Nascimbeni2022}. PLATO's mission target is to observe at least 245,000 FGK dwarf stars with a magnitude $V < 13$, 15,000 of which are high priority targets with a magnitude V < 11, as well as a further 5,000 M dwarfs with $V<16$ \citep{Montalto2021}. 

In this study, we aim to assess the capabilities of PLATO to study exoplanet populations across Galactic environments. We dynamically classified the likely target stars of the two proposed PLATO long-duration observation fields into three categories -- their thin disk, thick disk, and halo membership -- based on the \textit{Gaia} survey. Using the New Generation Planetary Population Synthesis (NGPPS) dataset, based on the Bern planet formation model \citep{Emsenhuber2021, Emsenhuber2021a}, we simulated the planetary populations in each Galactic component. We then estimated the detection efficiency of PLATO for these planet populations as a function of both the instrumental, planetary, and stellar properties, based on a model by \citet{Borner2022}, assuming a nominal PLATO pointing time of 2 years per field.

The structure of this paper is as follows. In Sect. \ref{sec:stellar_populations} we describe the selection of our stellar sample, and the classification of stars into their respective Galactic components. In Sect. \ref{sec:planet_populations} we detail the PLATO detection efficiency model, the NGPPS planet population dataset, and the methodology for creating mock planet populations and observations. In Sect. \ref{sec:results} we present the results of our analysis, including the detection efficiencies, planet occurrence rates, and expected PLATO planet yields for the two long-observation fields. Finally, in Sect. \ref{sec:discussion} we discuss the implications of our findings for understanding exoplanet demographics and the impact of Galactic environment and stellar properties on planet formation. A summary of our findings can be found in Sect. \ref{sec:conclusions}.

\section{Stellar populations}
\label{sec:stellar_populations}
\subsection{Sample selection and characterisation}
\label{subsec:stellar_sample}
The starting point for our analysis is the all-sky PLATO input catalogue (asPIC 1.1; \citealt{Montalto2021}). The aim of the PLATO mission is the search for terrestrial exoplanets in the habitable zone around solar-type stars, and the asPIC catalogue encompasses the most promising targets for this is goal, based on the \textit{Gaia} DR2 catalogue \citep{GaiaCollaboration2016, GaiaCollaboration2018}. They have been selected to lie within a specific region of the \textit{Gaia} colour-magnitude diagram to ensure they are mostly main-sequence FGK stars, with an additional smaller sample optimised for M-type dwarfs. A magnitude limit is imposed, optimised to maximise the probability of detecting terrestrial planets in the habitable zone. \citet{Montalto2021} also verified that this selection process does not introduce any bias in the metallicity distribution of the sample.

More specifically, the catalogue encompasses 2,675,539 stars, including 2,378,177 FGK dwarfs and subgiants brighter than a visual apparent magnitude of $V<13$ with a median distance of 428 pc, and 297,362 M dwarfs brighter than $V<16$ with a median distance of 146 pc. \citet{Montalto2021} determine the stellar fundamental parameters (effective temperature $T_\mathrm{eff}$, radius $R$, and mass $M$) for all target stars, and estimate overall uncertainties on these parameters to be 4\% for the $T_\mathrm{eff}$, 9\% for $R$ and 11\% for $M$. In our analysis, we restrict ourselves to the FGK sub-sample, requiring non-zero estimates for both mass, radius and effective temperature.

We cross-match these stars with the \textit{Gaia} DR3 catalogue using the cross-match table provided by the \citep{GaiaCollaboration2023} and retrieve parallaxes, proper motions, and radial velocities \citep{Lindegren2021, Seabroke2021}. Stars with uncertainties exceeding 20\% in any of these parameters are removed.

For this sample, we obtain metallicity [Fe/H] and $\alpha$-element abundance [$\alpha$/Fe] from the \textit{Gaia} DR3 medium-resolution RVS (Radial Velocity Spectrometer) spectra, using the best quality sample as defined by \citet{Recio-Blanco2023}\footnote{Specifically, this means that first 13 GSP-Spec quality flags for the RVS spectra are equal to zero. For more information, see Appendix C in \citet{Recio-Blanco2023}.}. This covers about 15\% of the sample. If RVS estimates are unavailable, the [Fe/H] values are filled in using the machine learning-based estimates by \citet{Andrae2023}, which make use of the low-resolution XP spectra published with \textit{Gaia} DR3, if possible. We do the same for the surface gravity $\log g$, and remove stars without a metallicity or surface gravity estimate. After these steps, 1,830,288 stars remain (68.4\% of the original sample).

We supplemented this information with [Fe/H], $\alpha$-element abundance and 
$\log g$ estimates from high-resolution spectra found in the APOGEE-DR17 
\citep{Abdurrouf2022} and GALAH DR3 \citep{Buder2021} spectroscopic survey databases, if available. We applied the following quality selection: For APOGEE, we 
require that none of the \texttt{STAR\_WARN}, \texttt{M\_H\_WARN}, or 
\texttt{ALPHA\_M\_WARN} quality flags are set. For GALAH, we do the same with the 
\texttt{flag\_sp}, \texttt{flag\_fe\_h}, and \texttt{flag\_alpha\_fe} flags. 
This allows us to retrieve information for a sub-sample of 73,632 target 
stars from APOGEE and 14,347 stars from GALAH. A discussion on the 
difference between the APOGEE/GALAH estimates and the \textit{Gaia} values for 
these properties can be found in \cref{appendixa:gaia_high_res_comparison}.

\begin{table}[ht]
    \centering
    \caption{Number of stars per stellar population in the all-sky sample, and in the southern (LOPS2) and northern (LOPN1) long-observation fields.}
    \begin{tabular}{l|r|r|r}
         Population & All-Sky & LOPS2 & LOPN1 \\
        \addlinespace
        \hline \hline
        \addlinespace \addlinespace
        Thin Disk & 1,160,029 & 80,954 & 85,626 \\
        Thick Disk Candidate & 599,549 & 38,534 & 41,585 \\
        Thick Disk & 64,103 & 3,988 & 4,415 \\
        Halo Candidate & 2,875 & 185 & 212 \\
        Halo & 3,732 & 241 & 240 \\
        \addlinespace
        Total & 1,830,288 & 123,902 & 132,078 \\
    \end{tabular}
    \label{tab:star_count_per_component}
\end{table}

\subsection{Galactic component classification}\label{subsec:component_classification}
To classify the stellar sample into three Galactic components: thin disk, thick disk, and halo, we follow the probabilistic approach outlined by \citet{Bensby2003a, Bensby2014c}, which has previously been applied in the context of exoplanet host star classification \citep{Bashi2020, Chen2021a, Bashi2022}. This method relies solely on stellar kinematics, specifically the Galactic velocities ($U_\mathrm{LSR}$, $V_\mathrm{LSR}$, $W_\mathrm{LSR}$) in the Local Standard of Rest (LSR). We assume that the Galactic velocities in the LSR follow a multivariate normal distribution, characterised by the probability density function:
\begin{equation}
    f_\mathrm{C}(U_\mathrm{LSR}, V_\mathrm{LSR}, W_\mathrm{LSR}) 
        = \mathcal{N} \left( 
        \begin{bmatrix}
           0 \\
           V_\mathrm{asym} \\
           0
         \end{bmatrix},
        \begin{bmatrix}
           \sigma_U \\
           \sigma_V \\
           \sigma_W
         \end{bmatrix}
        \right),
\end{equation}
where $f_\mathrm{C}$ represents the probability density of the star belonging to component $C$. The kinematic parameter $V_\mathrm{asym}$, $\sigma_U$, $\sigma_V$, and $\sigma_W$ differ for each Galactic component, and also depend on the star's Galactocentric radius ($R$) and height ($Z$). For this study, we utilise the revised kinematic characteristics presented by \citet{Chen2021b}.

To perform the classification, we compute the relative Galactic component membership probabilities, TD/D (thick disk-to-thin disk) and TD/H (thick disk-to-halo), for each star:
\begin{equation}
    \frac{\mathrm{TD}}{D} = \frac{X_\mathrm{TD}}{X_\mathrm{D}} \cdot \frac{f_\mathrm{TD}}{f_\mathrm{D}},
    \quad
    \frac{\mathrm{TD}}{H} = \frac{X_\mathrm{TD}}{X_\mathrm{H}} \cdot \frac{f_\mathrm{TD}}{f_\mathrm{H}},
\end{equation}
where $X$ represents the fraction of stars belonging to a given component and is similarly dependent on $R$ and $Z$.

Based on the membership probabilities, we categorise the stars into one of three components \citep{Bensby2014}: (\textit{i}) Thin disk star if TD/D < 0.1 (i.e. the thin disk membership probability is ten times larger than the thick disk membership probability), (\textit{ii}) Thick disk star if TD/D > 10 and TD/H > 10, and (\textit{iii}) Halo star if TD/H < 0.1. We apply these large probability ratios in order to reduce contamination. Values that fall between these limits are considered to be not clearly classifiable. They are labelled thick disk candidate (if 0.1 < TD/D < 10) and halo candidate (if 0.1 < TD/H < 10). About 1/3 of stars fall into these unclear candidate categories and will not be considered in our exoplanet population analysis. 

\begin{figure}
    \centering
    \includegraphics[width=\columnwidth]{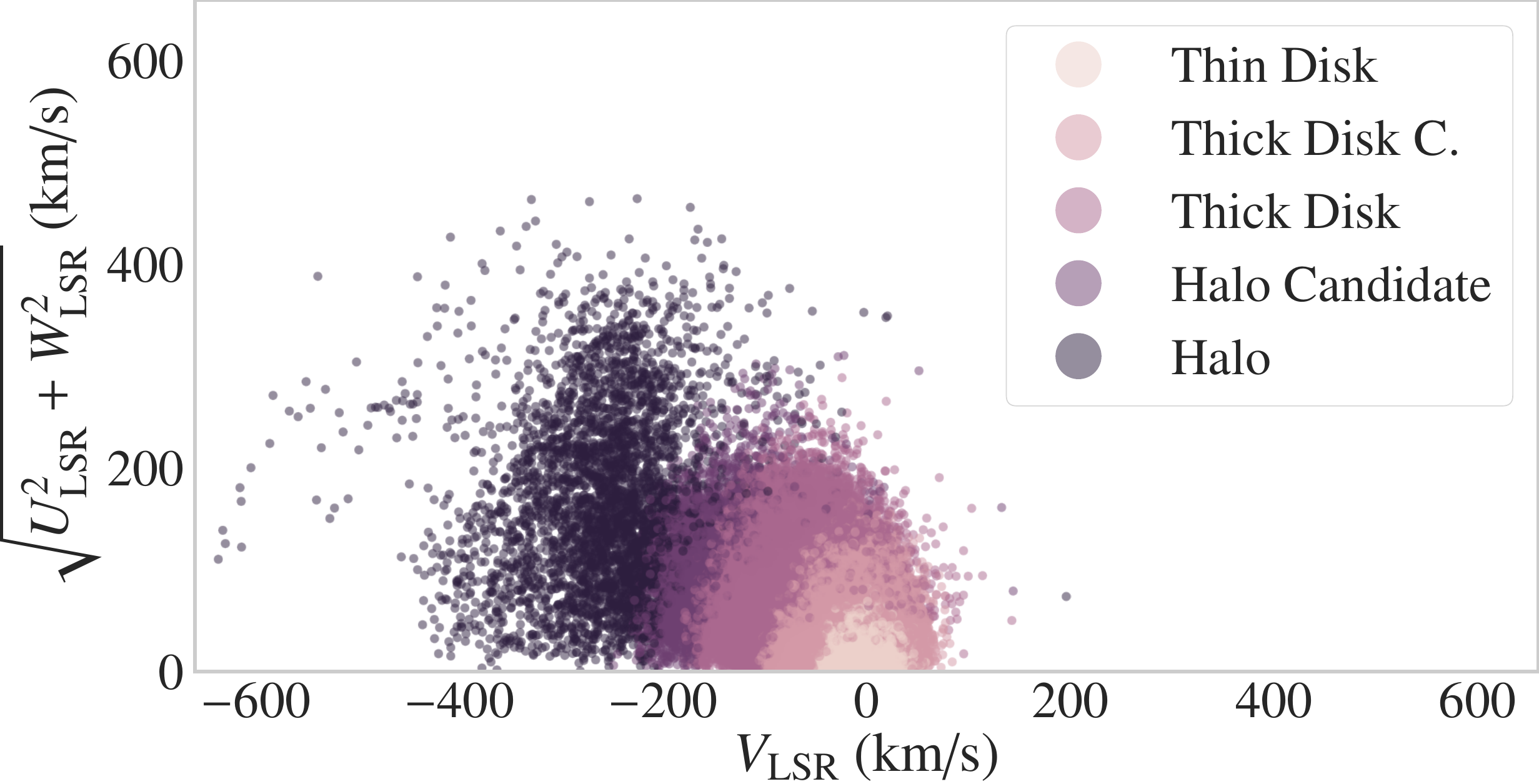}
    \caption{Toomre diagram of the all-sky sample. The stars are coloured according to their classification into the different Galactic components, as described in Sect. \ref{subsec:component_classification}.}
    \label{fig:toomre_scatterplot}
\end{figure}

The Galactic velocities and coordinates are calculated using the \texttt{galpy} python package \citep{Bovy2015}. The corresponding Toomre diagram is shown in Fig. \ref{fig:toomre_scatterplot}. For the all-sky sample, about 63.4\% of stars are classified as thin disk stars, and 36.2\% are classified as either thick disk candidates or thick disk stars. The halo candidate and halo star sub-sample makes up about 0.4\% of stars, totalling about 6,900 stars out of the approximately 2 million. A detailed breakdown can be found in Table \ref{tab:star_count_per_component}.

\subsection{Chemical abundances: [Fe/H] and [\texorpdfstring{$\alpha$}\ /Fe] distributions}
\label{subsec:metallicity}
Different Galactic components exhibit distinct stellar populations, characterised not only by kinematic variations \citep{Gilmore1983} but also by chemical composition differences \citep{Bensby2014c, Masseron2015, Hawkins2015}. These differences have the potential to significantly impact the exoplanet populations associated with each component \citep{Bashi2022, Nielsen2023, Cabral2023, Boettner2024}. Ideally, classifying stars into specific Galactic components should consider both kinematic and chemical information, as relying on a single method can result in significant contamination \citep{Bensby2014c, Bashi2019}.

Since only about 5\% of the sample have [Fe/H] and [$\alpha$/Fe] estimates from APOGEE or GALAH high-resolution spectra, we cannot uniformly apply this additional condition in our classification across the full sample. However, we can utilise this subset to confirm the effectiveness of the kinematic classification, and loosely quantify contamination.

\begin{figure}
    \centering
    \includegraphics[width=\columnwidth]{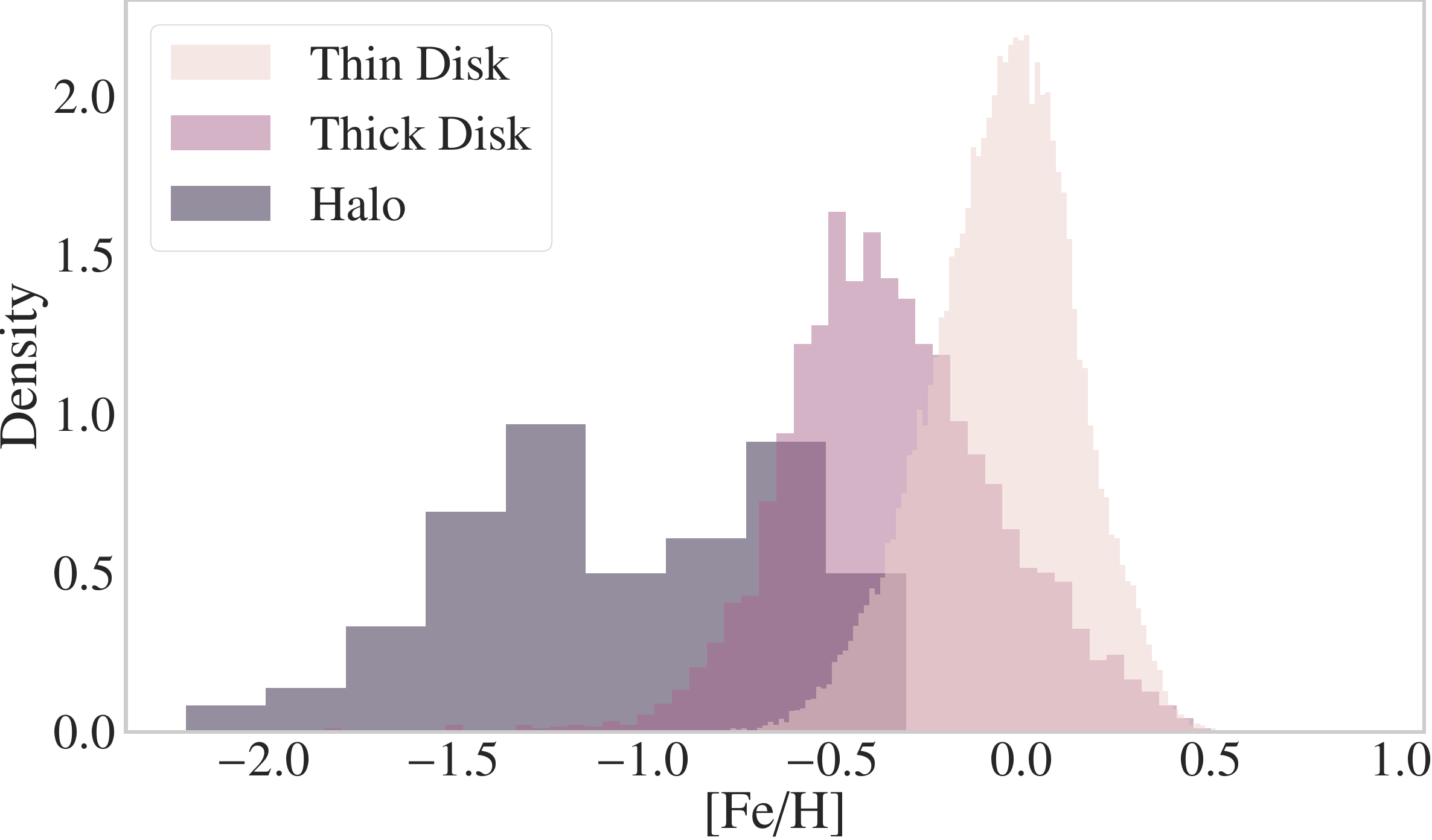}
    \caption{Histogram of the [Fe/H] metallicity distribution for the thin disk, thick disk, and halo components for the all-sky sample. Shown are only those stars that have reliable estimates from APOGEE or GALAH.}
    \label{fig:metallicity_hist}
\end{figure}

In terms of metallicity, halo stars are the most metal-poor, followed by thick disk and thin disk stars \citep{Bland-Hawthorn2016}. This trend is also observed in our kinematically classified sample. In Figure \ref{fig:metallicity_hist}, we show the metallicity distribution of the stars with [Fe/H] estimates available from APOGEE 
or GALAH. The thin disk population peaks around [Fe/H] = 0, the thick disk 
population around [Fe/H] =-0.4. The halo sub-sample only contains 176 stars, but has a clear bimodal distribution, with peaks at [Fe/H]= -0.7 and [Fe/H]= -1.2. The lower metallicity population constitutes primarily accreted stars, while the higher metallicity population are  stars formed in situ and expelled onto halo-like orbits by the last major merger \citep{Helmi2018, DiMatteo2019}.

\begin{figure}[ht]
     \centering
     \begin{subfigure}{\columnwidth}
        \centering
        \includegraphics[width=\textwidth]{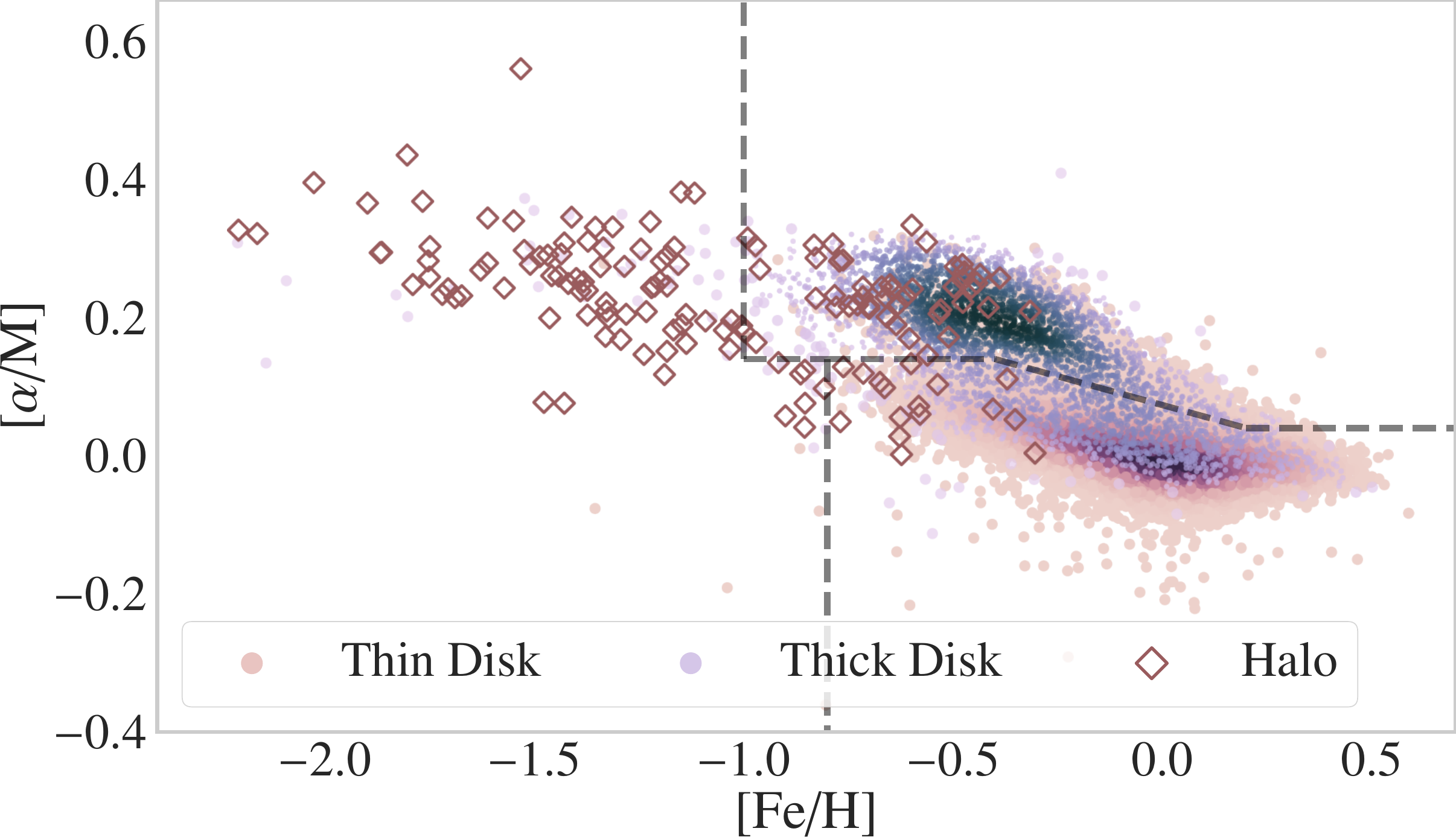}
        \caption{}
        \label{fig:metallicity_alpha_apogee}
     \end{subfigure}
     \begin{subfigure}{\columnwidth}
        \centering
        \includegraphics[width=\textwidth]{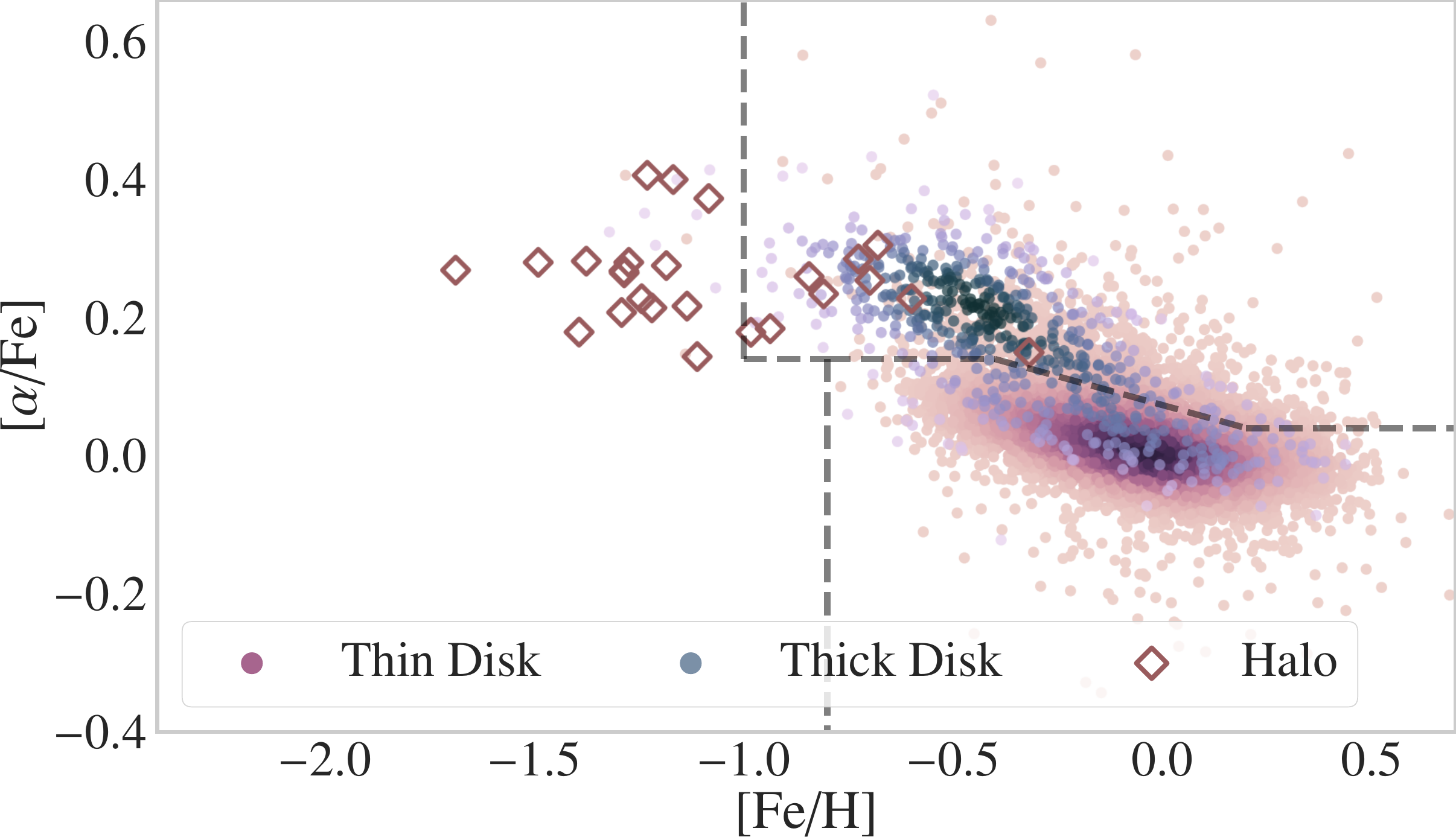}
        \caption{}
        \label{fig:metallicity_alpha_galah}
     \end{subfigure}
     \caption{Scatter plot of the all-sky sample in the metallicity - $\alpha$ element plane, categorised by their classification into the thin disk, thick disk, and halo component. Only those stars with corresponding APOGEE or GALAH estimates are shown. Panel (a): APOGEE. Panel (b): GALAH. Darker shades represent a higher density of points. Halo stars are marked as diamonds. Note that for the APOGEE sample, we show [$\alpha$/M], which may slightly differ from [$\alpha$/Fe].  The (dashed) chemical separation lines between halo (left section), thick disk (upper right section) and thin disk (lower right section) are based on \citet{Horta2023}.}
     \label{fig:metallicity_alpha}
\end{figure}

There is a large overlap in terms of metallicity between the thin disk and thick 
disk. Because of that, their populations are better distinguished through their 
distribution in the [Fe/H]--[$\alpha$/Fe] plane \citep{Prochaska2000, Reddy2003, Reddy2006}, which is well-known to be bimodal \citep{Wallerstein1962}. We show the 
distribution of the APOGEE and GALAH sub-samples in this plane in Fig. \ref{fig:metallicity_alpha}. The kinematically classified stars fall into three 
distinct clusters, which are largely consistent with the thin disk/thick disk/halo 
expectation: high metallicity ([Fe/H] $\sim 0$) and low $\alpha$ ([$\alpha$/Fe] $\sim 0$) for the thin disk, lower metallicity ([Fe/H] $\sim~-0.4$) and high $\alpha$ ([$\alpha$/Fe] $\sim 0.2$) for the thick disk, and low metallicity  ([Fe/H] $\sim <0.5$) for halo stars (see also Table \ref{tab:high_res_metallicity_distribution}).

\begin{figure}
    \centering
    \includegraphics[width=\columnwidth]{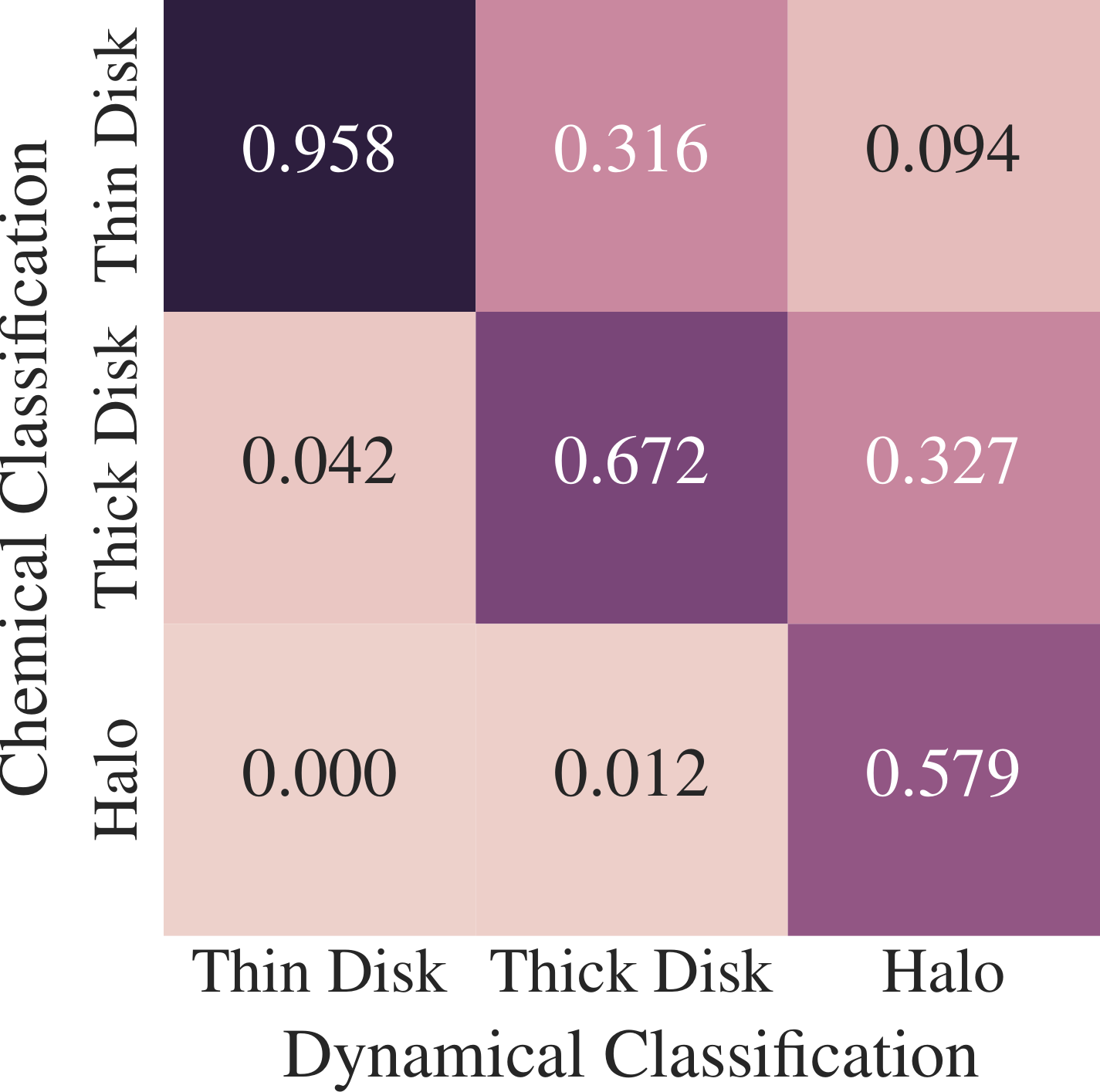}
    \caption{Confusion matrix comparing the dynamical and chemical classifications for stars possessing both metallicity and $\alpha$-abundance estimates from APOGEE or GALAH. Each tile presents the total number of stars (top), precision (middle, fraction normalised along columns), and recall (bottom, fraction normalised along rows). For instance, 3242 stars are consistently classified as belonging to the thick disk by both methods. This corresponds to 68.15\% of targets that were kinematically classified as thick disk stars, and 58.79\% of targets that were chemically classified as such.}
    \label{fig:confusion_matrix}
\end{figure}

To quantify the level of contamination between the kinematically defined populations, we independently classify all stars possessing both metallicity and $\alpha$-abundance estimates from high-resolution spectra into the three Galactic components. This chemical classification is based on the criteria defined by \citet{Horta2023}, as visualised in Fig. \ref{fig:metallicity_alpha}. The resulting confusion matrix, presented in Fig. \ref{fig:confusion_matrix}, reveals that 93.6\% of targets share consistent chemical and dynamical classifications. However, this high agreement is largely driven by the significant imbalance in the dataset towards thin disk stars, which are inherently more common.

More importantly, approximately 30\% of stars dynamically classified as belonging to the thick disk are chemically classified into the thin disk. This overlap is also visible in the metallicity distribution shown in Fig. \ref{fig:metallicity_hist}, where the thick disk population exhibits a more pronounced tail towards higher metallicity values ([Fe/H] > 0) than one would expect for a thick disk sample, indicative of thin disk contamination. Similarly, around 33\% of dynamically classified halo stars are chemically identified as thick disk stars, and a further 10\% as thin disk stars. However, these stars consistently have lower metallicities ([Fe/H] < -0.3) compared to the average metallicity of the thick disk and thin disk sample.

\begin{table}[ht]
    \centering
    \caption{Median metallicity and $\alpha$-element abundance from APOGEE-DR17 and GALAH DR3 for the different stellar populations; the values were calculated for the all-sky sample.}
    \begin{tabular}{l|cc|cc}
        \multirow{2}{*}{Population} & \multicolumn{2}{c}{APOGEE} & \multicolumn{2}{c}{GALAH} \\
        & [Fe/H] & [$\alpha$/M] & [Fe/H] & [$\alpha$/Fe] \\
        \hline \hline
        \addlinespace
        Thin Disk & -0.03 & 0.00 & -0.06 & 0.02 \\
        Thick Disk Candidate & -0.08 & 0.03 & -0.10 & 0.03 \\
        Thick Disk & -0.36 & 0.17 & -0.38 & 0.17 \\
        Halo Candidate & -0.65 & 0.24 & -0.77 & 0.26 \\ 
        Halo & -1.05 & 0.24 & -1.17 & 0.26 \\
    \end{tabular}
    \label{tab:high_res_metallicity_distribution}
    \tablefoot{For the APOGEE sample, we show [$\alpha$/M], which may slightly differ from [$\alpha$/Fe].}
\end{table}

\subsection{PLATO fields: LOPS2 and LOPN1}
\label{subsec:plato_fields}
The PLATO mission is designed to observe for at least 4 years in total. The observing strategy is yet to be finalised, but it is likely that PLATO will observe at least two fields during its Long-Duration Observation Phase (LOP). The minimum observation time for a LOP field is one year, with a planned time of at least two years \citep{Rauer2014a, Rauer2016}. For this study, we will assume a nominal observation strategy with two years being dedicated to the observation of the southern long-observation field (LOPS2), and a further two years to the provisional northern long-observation field (LOPN1).

PLATO's optical design includes 24 identical ``normal`` cameras (and an additional two ``fast`` with a higher cadence), grouped into arrangements of six, which will observe overlapping parts of the total field. This unique constellation results in a $49^\degree \times 49^\degree$ (approximately 5\% of the entire sky) square-shaped FoV, with parts being observed by 6, 12, 18, or 24 cameras \citep{Nascimbeni2022}. Additionally, these observations can be complemented by shorter "Step and stare" observations of about three months each, which could increase the coverage to 40\% of the sky.

The first long duration field LOPS2, located in the Southern Hemisphere, has recently been selected. It is located at $l=255.9375^\degree$ and $b=-24.62432^\degree$ in Galactic coordinates. The preliminary northern field LOPN1 is located at l=$81.56250^\circ$ and b=$24.62432^\circ$ \citep{Nascimbeni2022}. The northern field encompasses the entire \textit{Kepler} field and north TESS Continuous Viewing Zone (CVZ), whereas the southern field covers a large fraction of the south TESS CVZ.

\begin{figure}
    \centering
    \includegraphics[width=\columnwidth]{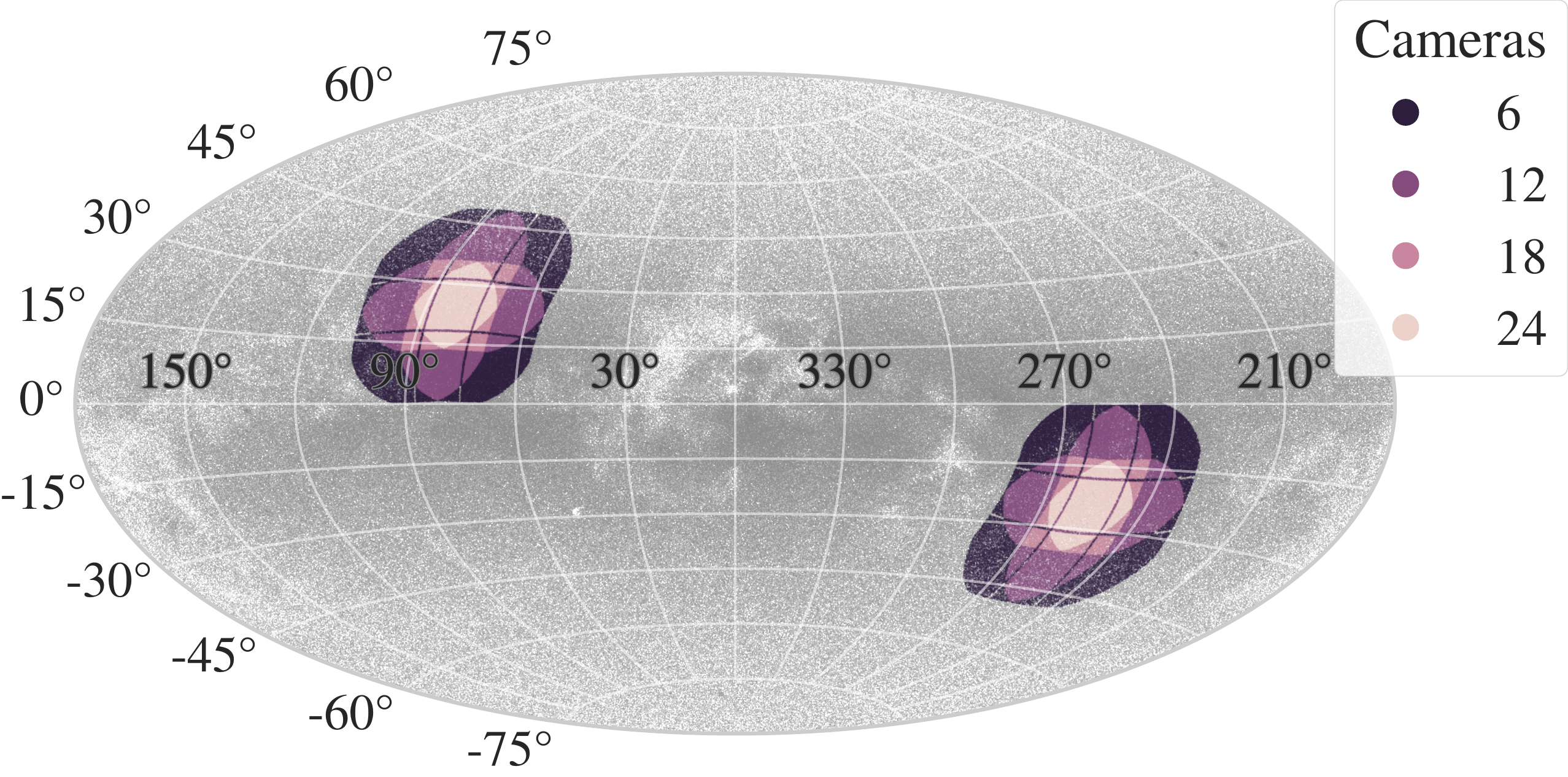}
    \caption{Scatter plot of the all-sky sample in Galactic coordinates (Aitoff projection). Stars that fall into the southern (LOPS2) and northern (LOPN1) long-observation fields are colour-coded by the number of cameras observing the target.}
    \label{fig:field_plot}
\end{figure}

We use a modified version of the publicly available \texttt{platopoint} code\footnote{\url{https://github.com/hposborn/platopoint}} to select the targets in the two LOP fields from the all-sky catalogue. The location and geometry of the fields are shown in Fig. \ref{fig:field_plot}. Combining both fields, a total of 255,980 target stars are covered (14.0\% of our sample). Of those, around 165,000 stars are classified as thin disk stars, 8,000 as thick disk stars, and about 500 stars falling into the halo category (see also Table \ref{tab:star_count_per_component}).

\section{Detection and exoplanet population models}
\label{sec:planet_populations}
\subsection{PLATO detection efficiency model}
\label{subsec:detection_efficiency}
The efficiency of planet detection depends greatly on the instrument, the detection pipeline, and the properties of the stellar system, including stellar magnitude, orbital inclination, etc. A typical approach to assess detection efficiencies for similar instruments, like \textit{Kepler} and TESS, is to use simulated or observed light curves, inject an artificial transit, and measure the recovery rate for those artificial planets. For this purpose, light curves are usually pre-processed and de-trended, after which transit detection is performed using either the box least squares (BLS, \citealt{Kovacs2002}) or transit least squares (TLS, \citealt{Hippke2019}) algorithm.

For this study, we take a simplified approach similar to the one employed by \citet{Matuszewski2023b}. We model the detection efficiency (DE) as a function of the signal-to-noise ratio (S/N)  $\mathrm{DE} (\mathrm{S/N})$. We define the S/N as
\begin{equation}
\mathrm{S/N} = \frac{\delta}{\sigma} \sqrt{n_\mathrm{transit}}
\end{equation}
(see also \citealt{Boley2021}) under the assumption of purely white noise. Here, $\delta$ is the transit depth, $\sigma^2$ is the variance of the de-trended light curve excluding the transit, and $n_\mathrm{transit}$ is the number of in-transit data points.

The next step is to link the signal-to-noise ratio to the detection efficiency. In the case of \textit{Kepler}, \citet{Fressin2013} employed a linear function. This function assigns a detection efficiency of 0\% for S/N values below 5, a detection efficiency of 100\% for S/N values above 16, and a linear increase in between these two points. However, \citet{Christiansen2015} suggested that a Gamma distribution provides a better fit to the \textit{Kepler} stellar sample.

We thus assume that the detection efficiency $\mathrm{DE} (\mathrm{S/N})$ is given by a cumulative Gamma distribution,
\begin{equation}
\mathrm{DE}~(\mathrm{S/N}) = \frac{c}{b^a \Gamma(a)} \int^\mathrm{S/N}_0 t^{a-1} \exp{\left(-t/b\right)}~\mathrm{d}t,
\end{equation}
with parameters a = 30.87, b = 0.271, c = 0.940 found by \citet{Christiansen2017} for \textit{Kepler}, although the specific parameter for PLATO will likely differ from those values.

With this parameterisation, $\mathrm{DE\leq 0.01}$ for an $\mathrm{S/N \leq 5.29}$, and $\mathrm{DE \geq 0.90}$ for $\mathrm{S/N \geq 11.13}$. This is slightly more conservative than the assumption in the PLATO Red Book, which assumes a detection efficiency of virtually 100\% for S/N > 10, based on \citet{Jenkins1996b, Jenkins2010b} and  \citet{Fressin2013}.

\begin{figure}
    \centering
    \includegraphics[width=\columnwidth]{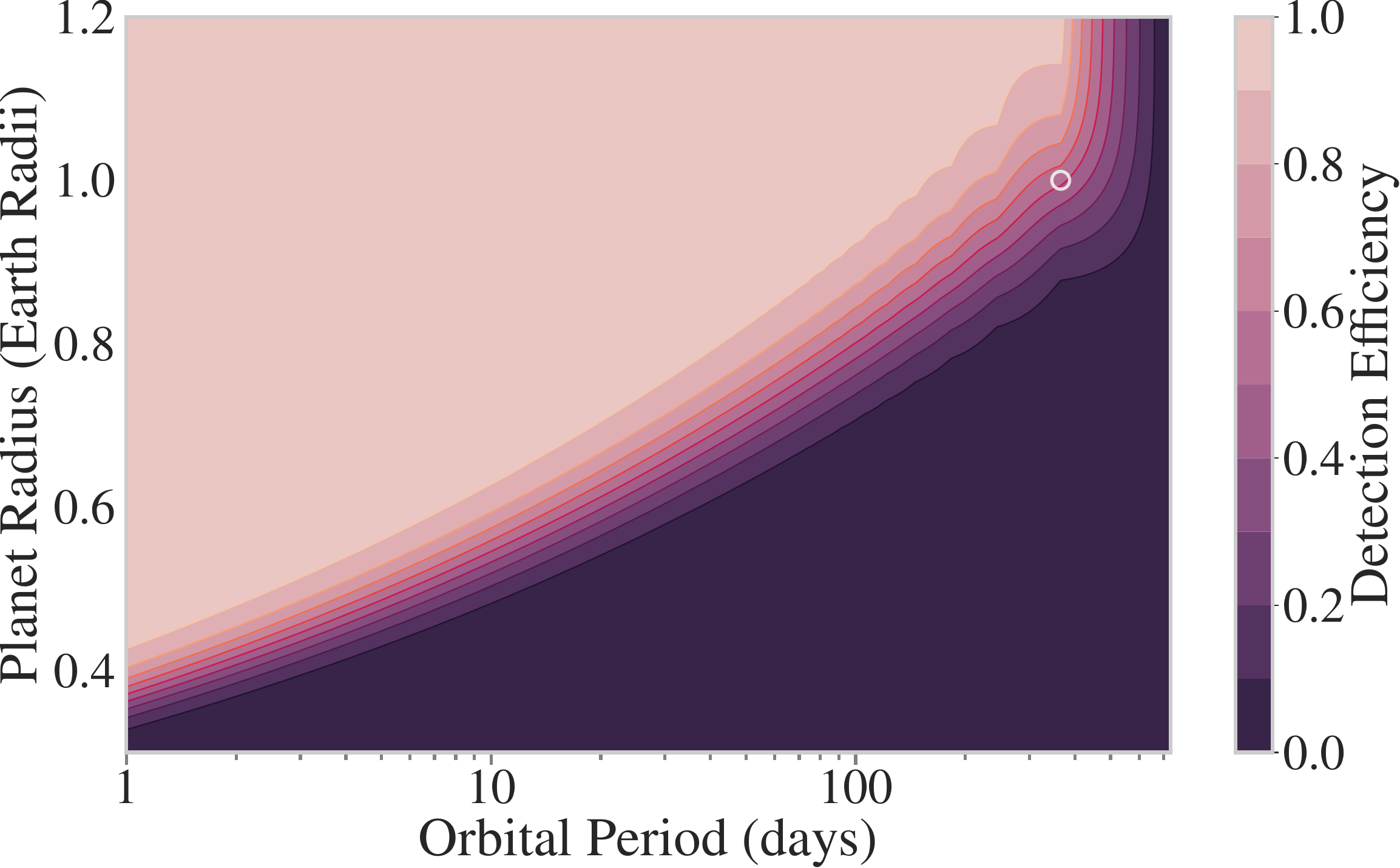}
    \caption{Contour plot of the detection efficiency as a function of $R_\mathrm{p}$ and $P$. The fixed parameters are $R_\mathrm{star} = R_\Sun$, $M_\mathrm{star} = M_\Sun$, $V = 11$, $n_\mathrm{cameras}=24$, $\cos i$ = 0, $\sigma_\mathrm{star} = 10^{-5}$, $u_1 = 0.4$ and $u_2=0.15$. The white marker is located at $P = 1$ year, $R = 1 R_\Earth$.}
    \label{fig:detection_efficiency_contour}
\end{figure}

We detail the definitions of $\delta$, $\sigma$ and $n_\mathrm{transit}$ in the next sections. The detection efficiencies for a variety of orbital periods and planet radii are shown in Fig. \ref{fig:detection_efficiency_contour}.

\subsection{Transit depth \texorpdfstring{$\delta$}{} }
\label{subsec:transitdepth}
We calculate the transit depth with a correction for limb darkening using the analytical treatment by \citet{Heller2019a}. The transit depth is given by the following equation:
\begin{equation}
\delta = \left(\frac{R_\mathrm{p}}{R_\mathrm{s}}\right)^2 \cdot \frac{I_b}{\langle I \rangle_A},
\end{equation}
where $R_\mathrm{p}$ and $R_\mathrm{s}$ are the radii of the planet and star, respectively. The impact parameter-dependent flux covered by the planet during mid-transit is denoted as $I_b$, while $\langle I \rangle_A$ represents the average intensity across the stellar disk.

Assuming a quadratic limb darkening law, \citet{Heller2019a} find
\begin{align}
        & I_b = 1 - u_1 \left( 1 - \sqrt{1-b^2} \right) - u_2  \left( 1 - \sqrt{1-b^2} \right)^2 \\
        & \langle I \rangle_A = 1 - \frac{u_1}{3} - \frac{u_2}{6},
\end{align}
where $b = a/R_\mathrm{s} \cos i$ is the impact parameter, $a$ is the semi-major axis of the planet's orbit, and $u_1$, $u_2$ are the limb darkening coefficients. The limb-darkening coefficients depend on $T_\mathrm{eff}$ and $\log g$, we estimate the values for our sample based on the grid provided for PLATO by \citet{Morello2022}.
your language preference (US or UK)

\subsection{Light curve variance \texorpdfstring{$\sigma^2$}{}}
\label{subsec:noise_model}
The primary metric to quantify the light curve variance $\sigma^2$ is the combined differential photometric precision (CDPP), which consists of two components: an instrument-related component and a stellar component. Again, assuming white noise, we can write $\sigma^2 = \sigma_\mathrm{instrument}^2 + \sigma_\mathrm{star}^2$.

To quantify the instrumental contribution, we implement the model described by \citet{Borner2022} for the PINE software, which was also for the PLATO field selection process \citep{Nascimbeni2022}.
In short, the noise budget is composed of three major components
\begin{equation}
        \sigma_\mathrm{instrument}^2 = \sigma_\mathrm{photon}^2 + \sigma_\mathrm{random}^2 + \sigma_\mathrm{systematic}^2,
\end{equation}
where $\sigma_\mathrm{photon}$ is the photon noise of the target, $ \sigma_\mathrm{random}$ are random noise components including readout and smearing noise, while $\sigma_\mathrm{systematic}$ describes systematic components including jitter and breathing noise. For a particular target, the instrumental noise is mostly influenced by the star's magnitude and the number of cameras observing it. 

For faint stars ($V > 12$), random noise is the primary noise component, while systematic noise becomes dominant for brighter stars ($V < 8$). In between these two regimes, photon noise dominates. According to the PLATO design specifications, the maximum random noise allowed for a $V = 11$ star in the end-of-lifetime (EOL) configuration observed with at least 22 cameras is 50 parts per million (ppm) for an integration time of one hour \citep{Borner2022}. This takes account of a failure of up to two cameras during the mission's lifetime.

The stellar component of the noise, $\sigma_\mathrm{star}$, is influenced by the inherent variability of the star due to factors like granulation, oscillations, and activity, as well as any residual uncertainties after the light curve de-trending process. Studies by \citet{Morris2020} and \citet{Heller2022} estimate the stellar variability for PLATO targets to be between 100 and 200 ppm in a one-hour window. Similar values have been found by \citet{Zhang2024} for the LAMOST-\textit{Kepler} common star sample. With a cadence of 25s per exposure for the normal cameras \citep{Borner2022}, this amounts to a stellar noise standard deviation between 8 and 17 ppm if the signal is integrated over one hour. These estimates align with those by \citet{Matuszewski2023b} and \citet{vanCleve2016}. For this study, we adopt a value of 10 ppm across all stars. The effects of stellar variability on the number of exoplanet detections is further elaborated on in \cref{appendixb:stellar_variability}.

\subsection{Transit observations \texorpdfstring{$n_\mathrm{transit}$}{}}
The number of in-transit data points is given by
\begin{equation}
n_{\mathrm{transit}} = N_{\mathrm{transit}} \cdot\frac{t_{\mathrm{transit}}}{t_{\mathrm{integration}}},
\end{equation}where $N_{\mathrm{transit}}$ is the number of observed transits, $t_{\mathrm{transit}}$ is the transit duration and $t_{\mathrm{integration}} = 1$ refers to the integration time specified for the noise model \citep{Borner2022}.

Assuming an eccentricity of zero, the number of guaranteed transits is simply given by
\begin{equation}
        N_\mathrm{transit} = \left\lfloor{\frac{t_\mathrm{point}}{P}}\right\rfloor.
\end{equation}
Here, $t_\mathrm{point}$ is the pointing duration of a particular field (nominal two years for the long observation fields) and $P$ the orbital period of the planet. The transit duration is
\begin{equation}
t_\mathrm{transit} =
\frac{P}{\pi} \arcsin \left( \frac{R_\mathrm{s}}{a} \left\{ \left[ 1 + \left( \frac{R_\mathrm{p}}{R_\mathrm{s}} \right) \right]^2
 - \left[ \left( \frac{a}{R_\mathrm{s}} \right) \cos i \right]^2 \right\}^{1/2} \right),
\end{equation}
where $R_mathrm{p}$ and $R_mathrm{s}$ are the radii of the planet and star, respectively; $a$ is the semi-major axis; and $i$ is the orbital inclination \citep{Seager2003a}.

Aside from the guaranteed number of transits ($N_\text{transit}$), an additional transit may be observed, depending on the timing of the initial transit. The overall detection efficiency of a target is calculated as a weighted average of the detection using $N\_\mathrm{transit}$ and $N\_\mathrm{transit}+1$ transits. The probability of observing an additional transit is
\begin{equation}
        p_\mathrm{N+1} = \frac{t_\mathrm{point}~(\mathrm{mod}~P)}{P}
\end{equation}

Finally, for a detection to be considered valid, at least two transits are required. Otherwise, the corresponding detection efficiency is set to zero.

\subsection{The NGPPS planet population dataset}
In order to study the planet populations in the different stellar populations, and estimate planet yields for PLATO, we employ the New Generation Planet Population Synthesis (NGPPS) dataset, based on the third generation Bern global model for planet formation and evolution \citep{Emsenhuber2021}. The model is based on the core-accretion paradigm, encompassing the growth of planets from 1 to 100 embryos (each about $0.01 M_\Earth$) into planets ranging from Earth-like to Super-Jupiter giants. The planet embryos grow in fluid-like protoplanetary disks described by gas and planetesimal fields. Direct N-body interactions between growing planets are also included, which increases the number of massive planets compared to isolated evolution.

The model simulations are initialised when planet embryos have formed, bypassing the dust-to-planet embryo growth stages. The number and distribution of embryos are free parameters. The formation phase lasts 20 Myr, after which planets evolve in isolation, influenced by tidal migration and atmospheric escape. This phase duration is longer than typical protoplanetary disk lifetimes, encompassing most growth stages but not fully simulating the late-stage giant impact phase, potentially underestimating rocky planet masses. \citet{Emsenhuber2021} found planet formation mostly completes within $\sim 1$ AU after 20 Myr, while more distant planets remain underdeveloped.

A wide range of physical processes and parameters are included in the model. Among the varied parameters, the initial number of embryos and stellar metallicity are crucial. The number of embryos ($N_\mathrm{Embryo}$) directly influences the final number and properties of the planetary systems, with \citet{Emsenhuber2021a} studying scenarios with 1, 10, 20, 50, and 100 embryos. For consistency, we focus on the 10-100 embryos range. Stellar metallicity is directly related to the dust-to-gas ratio, critical for planet formation.

\citet{Emsenhuber2021} and \citet{Burn2021} have performed extensive population runs, classifying planets into categories based on their final masses. The host star mass is fixed at $1 M_\Sun$ in \citet{Emsenhuber2021a}, while \citet{Burn2021} extend the analysis to stars with 10\%, 30\%, 50\%, and 70\% of the Sun's mass, finding systematic correlations between higher [Fe/H] and total disk mass with an increased number of giant planets.

In this study, we focus on FGK-type stars, and therefore, we utilise the $1 M_\Sun$ model. \citet{Emsenhuber2021a} conducted 1000 simulations for this model, varying stellar metallicity, among other parameters, for each initial embryo count. This population of 1000 systems serves as the foundation for populating our stellar sample with target planets, enabling us to investigate the planetary population.

\subsection{Creating mock populations and observations}
\label{subsec:mocks}
To create mock planet populations for our stellar sample, we follow these steps:

(i) Assign planetary systems: We randomly assign each star in the stellar sample a planetary system from the 1000 planetary systems in the NGPPS population (with $N_\mathrm{Embryo}$ fixed). Since planetary system architecture is strongly correlated with host star metallicity, we use an assignment probability
\begin{equation}
\log_{10} p = - \alpha \sqrt{\left([\mathrm{Fe/H}]_\mathrm{star} - [\mathrm{Fe/H}]_\mathrm{system}\right)^2},
\end{equation}
with $\alpha = 10$, so that each 0.1 dex difference in metallicity makes the assignment 10 times less likely. \citet{Emsenhuber2021a} initially assumed a Gaussian metallicity distribution (with mean $\mu = -0.02$ and standard deviation $\sigma = 0.22$), so we divide by this probability to avoid skewing our sample towards more common systems in the NGPPS dataset.

(ii) Assign system inclination: We assign each system a random inclination by drawing a sample of $\cos i$ from a uniform distribution [0, 1].

(iii) Roche limit check: For consistency, we remove any planet whose semi-major axis falls within the Roche limit
\begin{equation}
d_\mathrm{Roche} = 2.44 R_\mathrm{star} \left( \frac{\rho_\mathrm{star}}{\rho_\mathrm{planet}} \right)^{1/3}.
\end{equation}

(iv) Metallicity limit (optional): Since NGPPS is restricted to a metallicity range of [-0.6, 0.5], most low-metallicity stars will be assigned systems with metallicity around -0.6. Alternatively, following \citet{Andama2024}'s suggestion that stars with [Fe/H] < -0.6 are unlikely to form planetesimals, we can remove planets from such systems.

To create mock observations from these mock populations, we calculate the detection efficiency for each planet and randomly select a sub-sample based on those probabilities. We assume an eccentricity of zero for all orbits, and assign a detection efficiency of zero to non-transiting planets.

\section{Results}
\label{sec:results}
\begin{figure}[t]
     \centering
     \begin{subfigure}{\columnwidth}
        \centering
        \includegraphics[width=\textwidth]{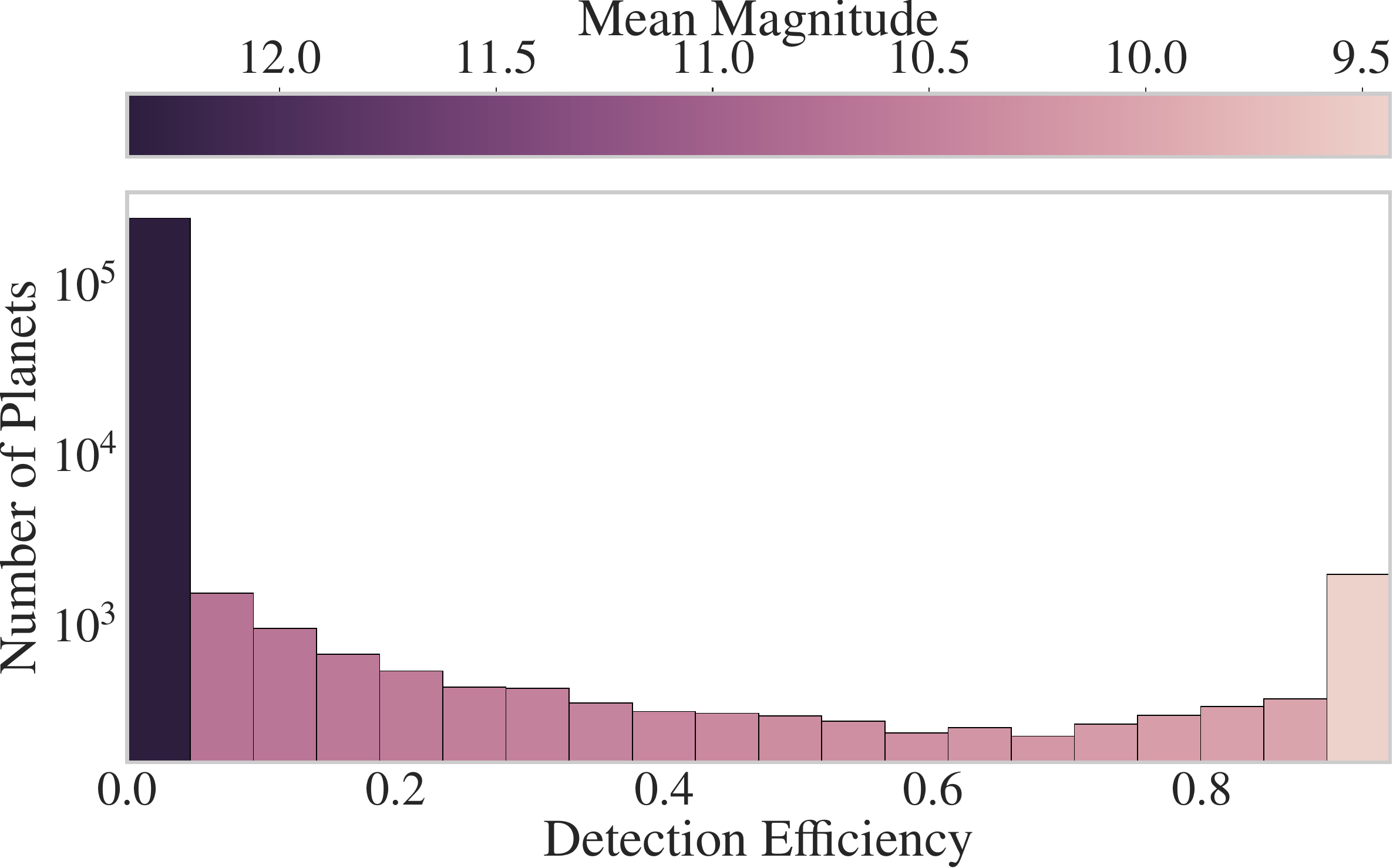}
        \caption{}
        \label{fig:Earthlike_detection_eff_hist}
     \end{subfigure}\vspace{2mm}
     \begin{subfigure}{\columnwidth}
        \centering
        \includegraphics[width=\textwidth]{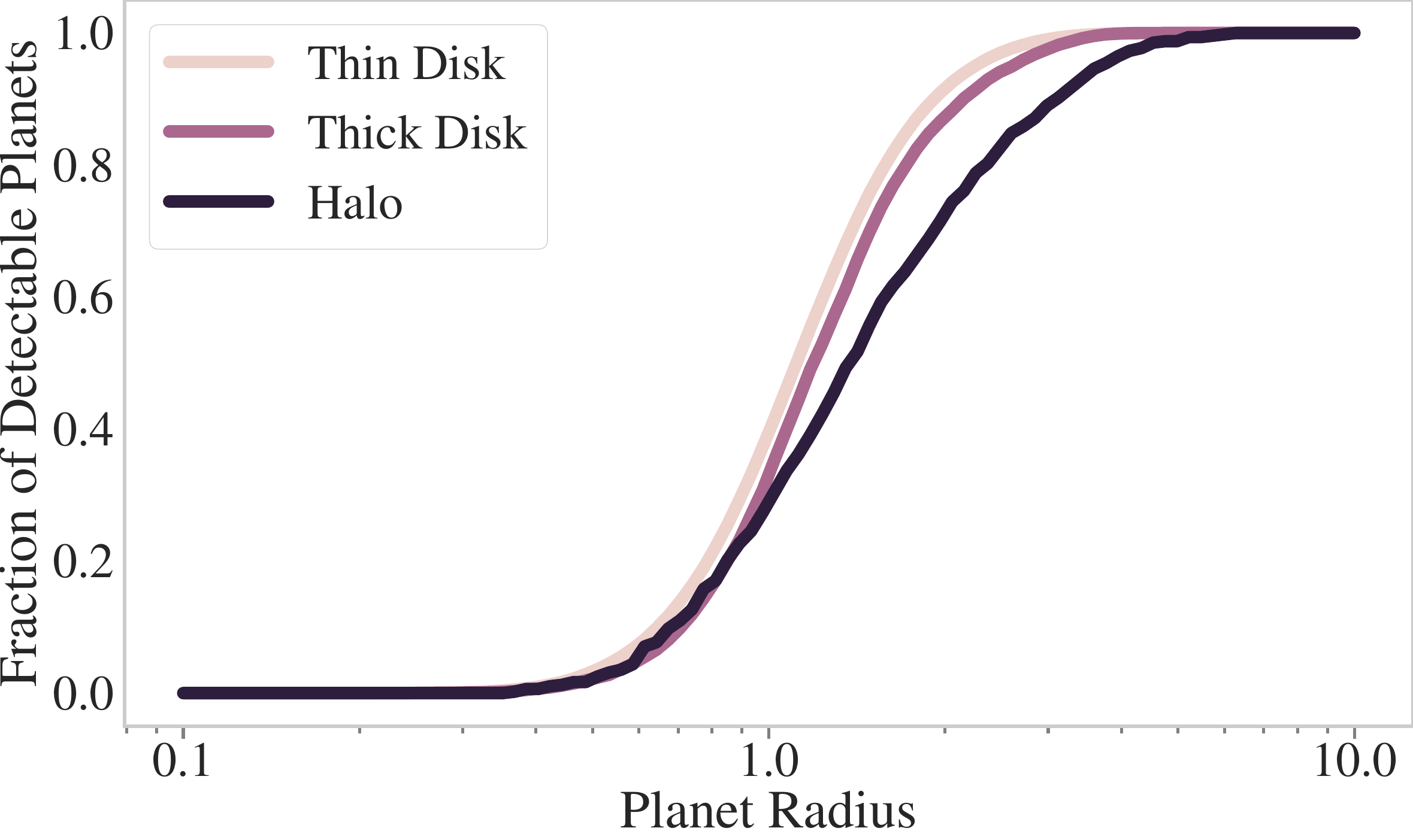}
        \caption{}
        \label{fig:detectioneff_planet_radius}
     \end{subfigure}
     \caption{Detection efficiency for the LOPS2 + LOPN1 targets. (a) The distribution of detection efficiencies calculated for the targets, assuming the transit of an Earth analogue ($P = 1$ year, $R = 1 R_\Earth$) with $\cos i = 0$. The colour of the bars encodes the mean magnitude of the stars within each bin. (b) The fraction of detectable planets (defined as planets with a detection efficiency $\geq 1/N_\mathrm{sample}$) for the thin disk, thick disk and halo populations, as a function of radius. The detection efficiencies are calculated for $\cos i=0$ and $P=100$ days.}
     \label{fig:detection_efficiency}
\end{figure}

\begin{figure}[t]
     \centering
     \begin{subfigure}{\columnwidth}
        \centering
        \includegraphics[width=\textwidth]{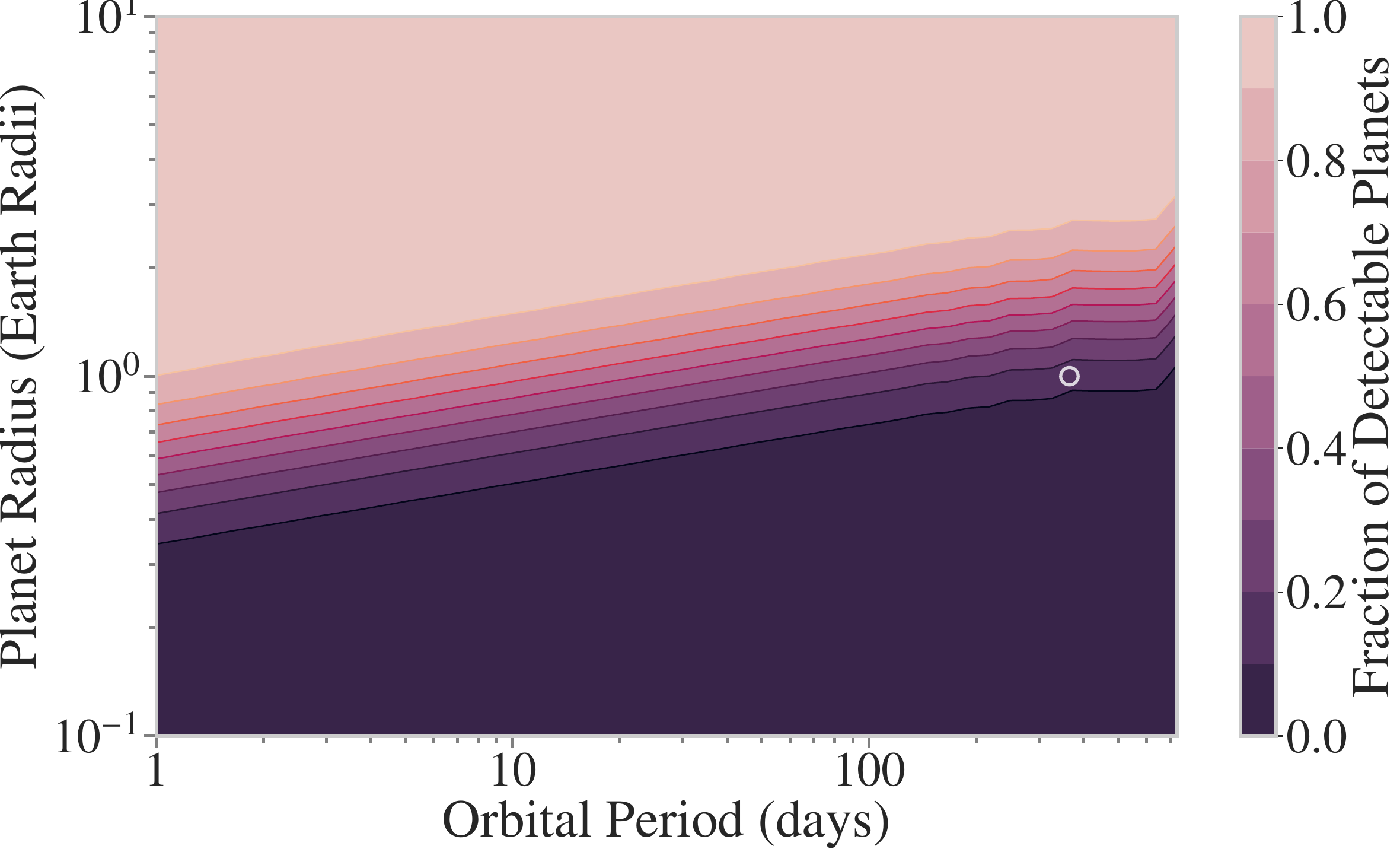}
        \caption{}
        \label{fig:detection_efficiency_fraction_all}
     \end{subfigure}\vspace{2mm}
     \begin{subfigure}{\columnwidth}
        \centering
        \includegraphics[width=\textwidth]{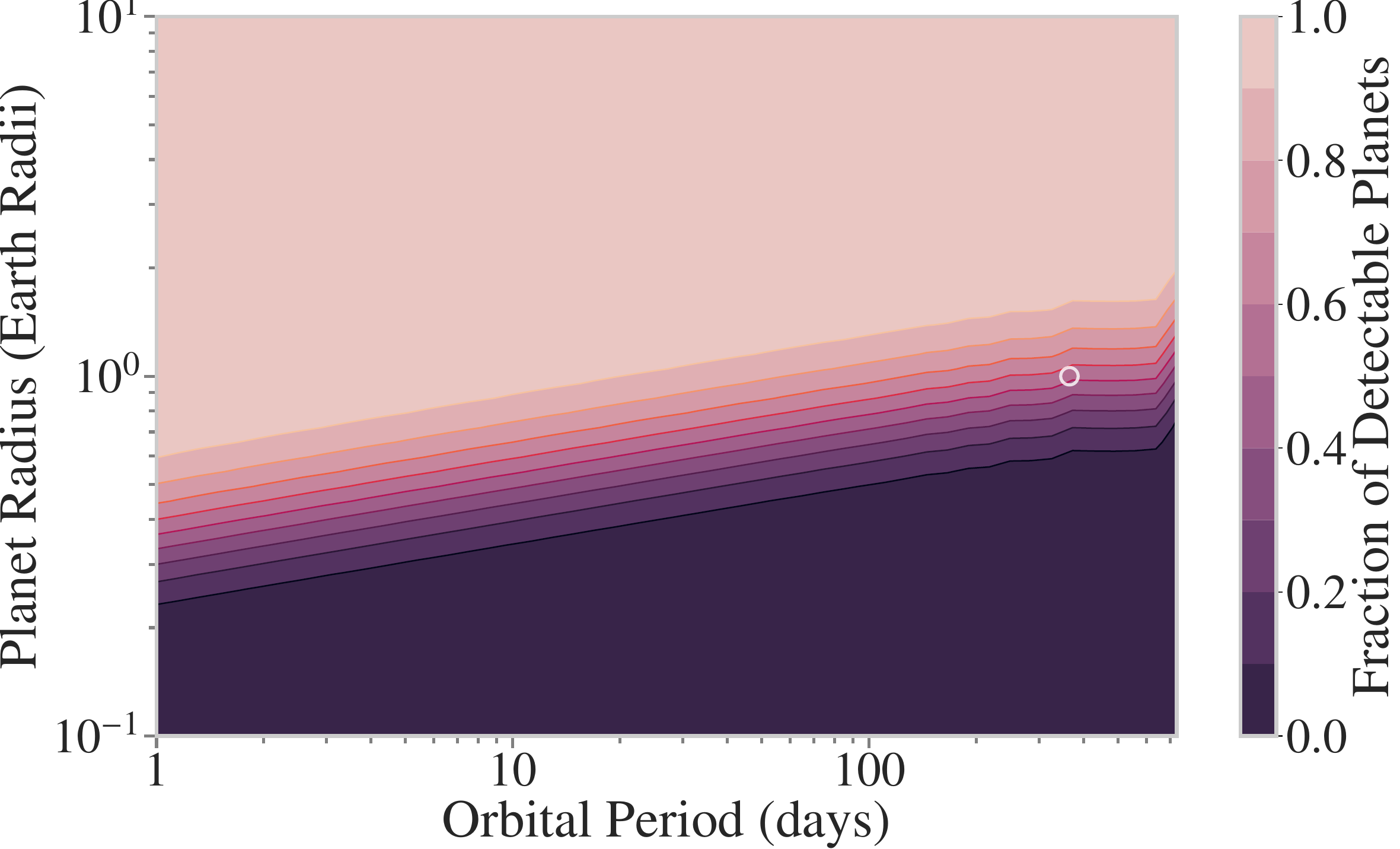}
        \caption{}
        \label{fig:detection_efficiency_fraction_p1}
     \end{subfigure}
     \caption{Fraction of detectable planets for the whole sample and P1 sub-sample, as a function of planet radius and orbital period. Detectable planets are defined as planets with a detection efficiency $\geq 1/N_\mathrm{sample}$, assuming $\cos i = 0$ for all planets. The white marker is located at $P = 1$ year, $R = 1 R_\Earth$. (a) All LOPS2 + LOPN1 targets. (b) P1 sample ($V \leq 11$ and random noise $\leq 50$ ppm integrated over one hour).}
     \label{fig:detection_efficiency_fraction}
\end{figure}

\subsection{Detection efficiency as a function of stellar and planetary properties}
\label{subsec:detection_eff_results}
For a fixed planet radius and orbital period, the detection efficiency is primarily determined by the stellar magnitude, stellar radius, and to a lesser extent, the number of cameras observing the target. The detection efficiency is higher for larger planets, shorter orbital periods, smaller stars, and brighter stars.

The distribution of detection efficiencies for all LOPS2 and LOPN1 targets, assuming each star hosts an Earth analogue (1 Earth radius, 1-year orbital period) with a transit aligned with the line of sight ($\cos i = 0$), is shown in Fig. \ref{fig:Earthlike_detection_eff_hist}. The distribution is strongly bimodal, with the vast majority of targets assigned a near-zero detection efficiency. Those stars are primarily faint ($V \geq 12$) or large ($R \geq 1.5 R_\Sun$). High detection efficiencies ($>0.9$), are found for bright ($V\leq 9.5$), small ($R<1R_\Sun$) stars. Intermediate values are relatively rare, since the detection efficiency rises from near-zero to almost one just by doubling the signal-to-noise ratio from 5 to 11 (see Sect. \ref{subsec:detection_efficiency}).

We define ``detectable planets`` as those with a detection efficiency of at least 1/$N_\mathrm{sample}$, where $N_\mathrm{sample}$ is the number of targets in the thin disk, thick disk, or stellar halo population to which the host belongs. This threshold guarantees that, on average, at least one planet would be found if each target star in the sample possessed a transiting planet with the same orbital period and radius. 

In Fig. \ref{fig:detectioneff_planet_radius}, we show the fraction of detectable planets as a function of planet radius for the thin disk, thick disk, and halo populations, assuming a 100-day orbital period and $\cos i = 0$. For smaller planets ($R \leq 1 R_\Earth$), the detection efficiency is nearly uniform across all populations. However, for larger planets, the halo population exhibits a smaller fraction of detectable planets compared to the thin and thick disk populations. This difference arises because the halo population contains a bigger proportion of larger or fainter stars, which decreases detectability for a planet of a given size.

Figure \ref{fig:detection_efficiency_fraction} displays the fraction of detectable planets as a function of both planet radius and orbital period for all LOPS2 and LOPN1 targets, as well as for the high signal-to-noise PLATO P1 sample. The P1 sample contains all targets with the additional requirements that $V \leq 11$ and that the random noise level is below 50ppm in one hour. Across the LOPS2 and LOPN1 fields, 29,238 target stars fulfil this requirement. For the entire sample, an Earth analogue would be detectable around 10–20 \% of stars. For the P1 sample, this number increases to 50\%.

\subsection{Planet occurrence rates and multiplicities across stellar populations}
\label{subsec:planet_populations}
\begin{table}
    \centering
    \caption{Planet categories as defined in \citet{Emsenhuber2021a}.}
    \begin{tabularx}{\columnwidth}{l|R|R}
         & Min. Mass $\left(M_\Earth\right)$ & Max. Mass $\left(M_\Earth\right)$ \\
        \addlinespace
        \hline\hline
        \addlinespace
        Earth & 0.5 & 2 \\ 
        Super-Earth & 2 & 10 \\ 
        Neptunian & 10 & 30 \\ 
        Sub-giant & 30 & 300 \\
        Giant & 300 &  \\ 
    \end{tabularx}
    \label{tab:planet_categories}
\end{table}

\begin{table*}
    \centering
    \caption{\label{tab:planet_metrics}Key planet population metrics, calculated for a single mock population with $N_\mathrm{Embryos} = 100$ across all targets in the LOPS2 and LOPN1 fields.}
    \begin{tabularx}{\textwidth}{ll|R|R|R|R}
     &  & Number of Planets & System Fraction & Occurrence Rate & Multiplicity \\
    \addlinespace
    \hline \hline
    \addlinespace
    \multirow{5}{*}{Thin Disk} & Earth & 759,616 & 0.91 & 4.56 & 5.02 \\
     & Super-Earth & 729,729 & 0.82 & 4.38 & 5.32 \\
     & Neptunian & 63,771 & 0.27 & 0.38 & 1.4 \\
     & Sub-Giant & 15,500 & 0.07 & 0.09 & 1.25 \\
     & Giant & 42,433 & 0.16 & 0.26 & 1.58 \\
     \addlinespace
    \hline
    \addlinespace
    \multirow{5}{*}{Thick Disk} & Earth & 45,396 & 0.90 & 5.40 & 6.03 \\
     & Super-Earth & 23,699 & 0.65 & 2.82 & 4.35 \\
     & Neptunian & 1,438 & 0.12 & 0.17 & 1.38 \\
     & Sub-Giant & 358 & 0.03 & 0.04 & 1.33 \\
     & Giant & 878 & 0.07 & 0.1 & 1.56 \\
     \addlinespace
    \hline
    \addlinespace
    \multirow{5}{*}{Halo} & Earth & 2,894 & 0.89 & 6.02 & 6.75 \\
     & Super-Earth & 393 & 0.40 & 0.82 & 2.06 \\
     & Neptunian & 5 & 0.01 & 0.01 & 1.00 \\
     & Sub-Giant & 1 & 0.00 & 0.00 & 1.00 \\
     & Giant & 2 & 0.00 & 0.00 & 2.00 \\
    \end{tabularx}    
    \tablefoot{The columns are the total number of planets, the fraction of systems with a planet of the given type, the occurrence rate (the total number of planets of this type divided by the total number of target stars), and multiplicity (the average number of planets of a given type around a star hosting at least one such planet).}
\end{table*}

To investigate the planet populations around different types of stars, we categorise the planets in our mock populations into five types based on their mass: Earth, super-Earth, Neptunian, sub-giant, and giant. This is the same classification as done by \citet{Emsenhuber2021}, and used in our previous work \citep{Boettner2024}. The exact definitions can be found in Table \ref{tab:planet_categories}. 

For each planet type, we calculate the key metrics: total number of planets, fraction of systems with a planet of the given type, occurrence rate (the total number of planets of this type divided by the total number of target stars), and multiplicity (the average number of planets of a given type around a star hosting at least one such planet). An example for a single mock population is shown in Table \ref{tab:planet_metrics}.

These metrics are calculated for each stellar population (thin disk, thick disk and halo) and for different assumptions about the initial number of embryos in the planet formation simulations (Table \ref{tab:number_of_planets}).

For the $N_\mathrm{Embryo} = 100$ simulations, we find that Earth-like planets are the most common across all stellar populations. The  occurrence rate and multiplicity increases with decreasing metallicity in the stellar populations, with an increase in the occurrence rate from 4.56 in the thin disk to 6.02 in the halo. More massive planets become a lot rarer (in comparison to Earth-like planets), as the average metallicity of the stellar population decreases. While in the thin disk, there are roughly as many Earth-like planets as there are super-Earths, they outnumber the super-Earths by a factor of seven in the halo. This trend becomes more extreme as the planet mass increases. For the halo population, there are predicted to be 1000 times as many Earth-like planets as there are giant planets.

\begin{figure}[t]
     \centering
     \begin{subfigure}{\columnwidth}
        \centering
        \includegraphics[width=\textwidth]{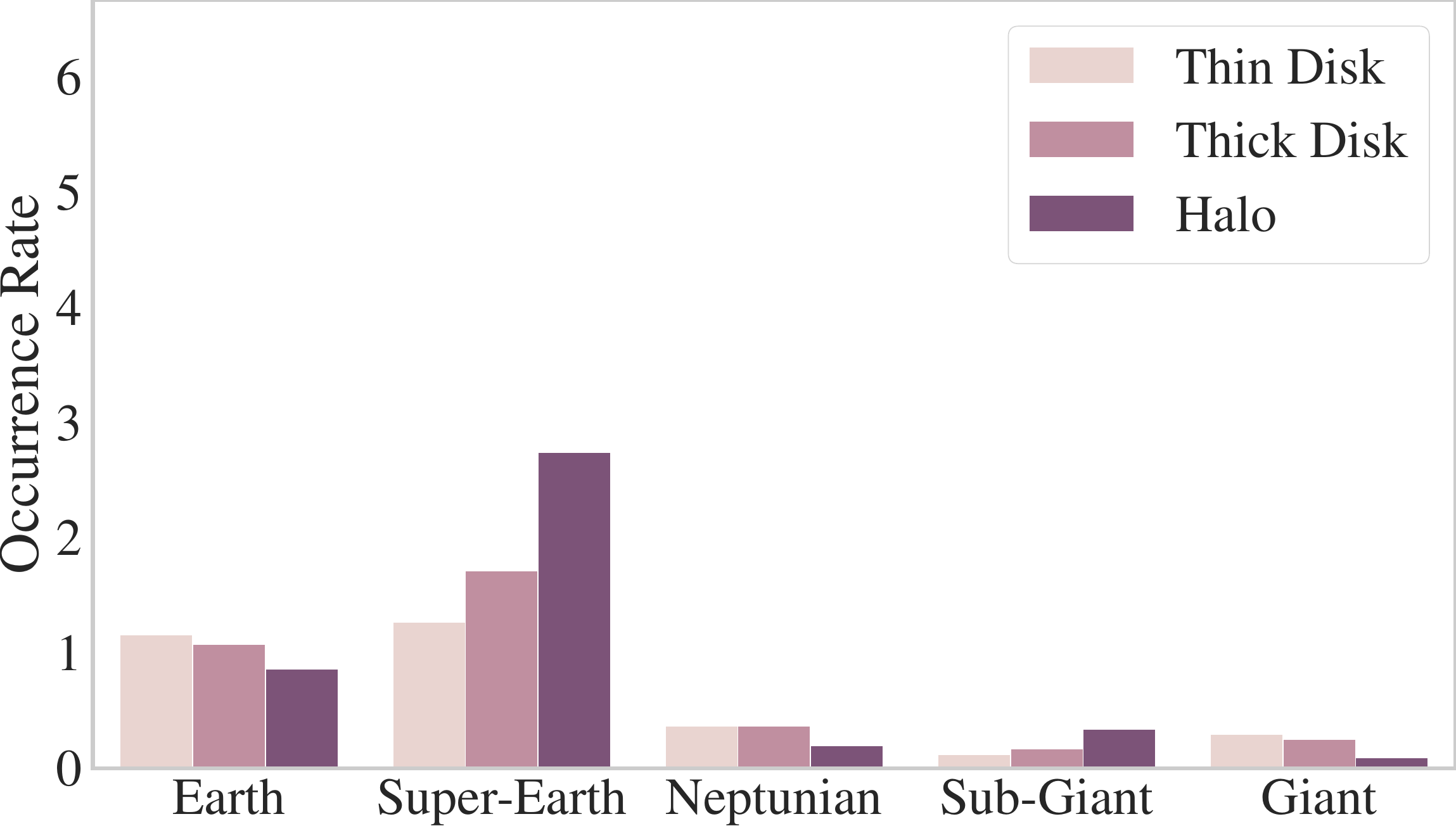}
        \caption{}
        \label{fig:occurence_rates_with_N_Embryos=10}
     \end{subfigure}\vspace{2mm}
     \begin{subfigure}{\columnwidth}
        \centering
        \includegraphics[width=\textwidth]{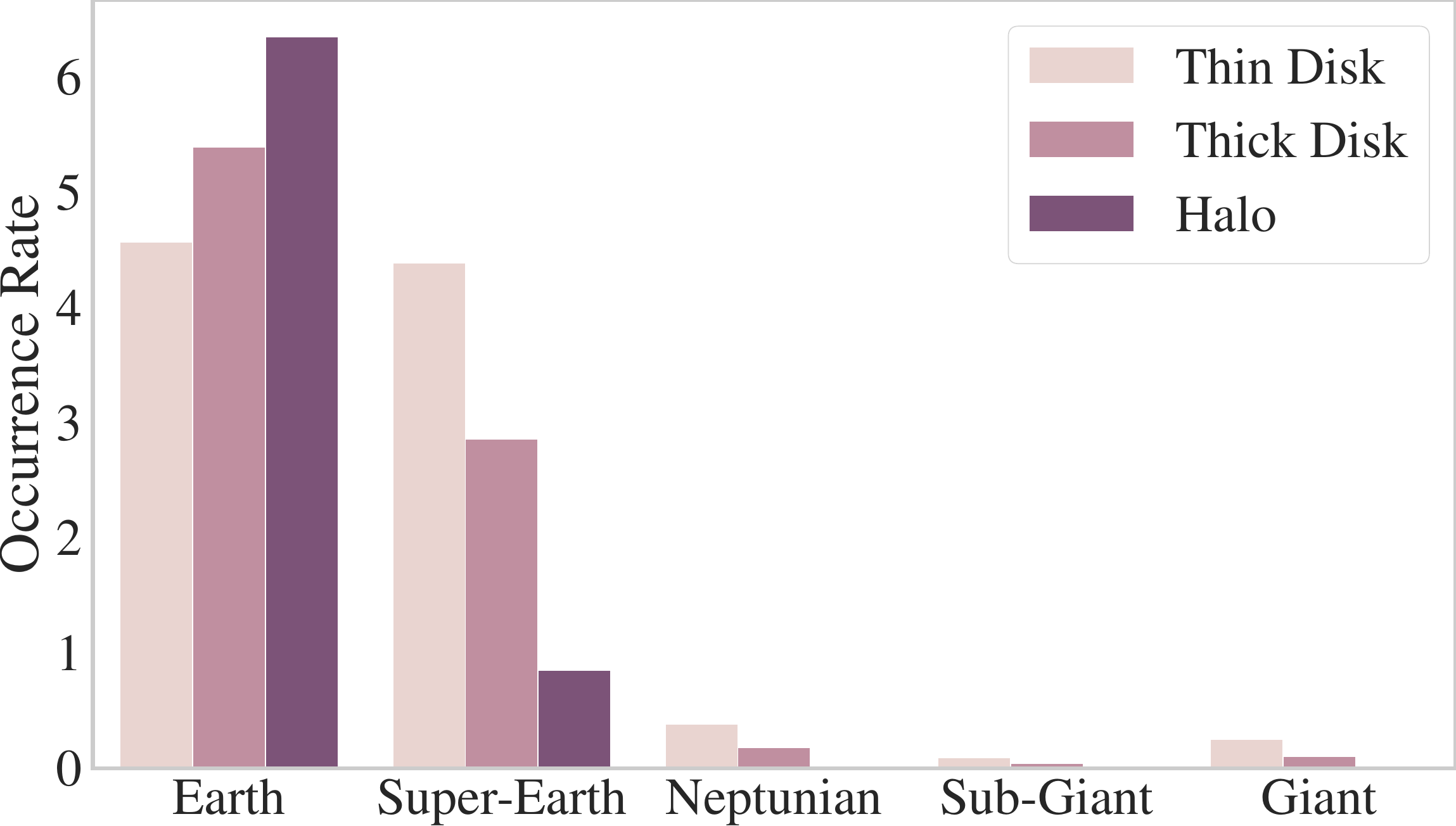}
        \caption{}
        \label{fig:occurence_rates_with_N_Embryos=100}
     \end{subfigure}
     \caption{Occurrence rates of different planet types per stellar population, for varying $N_\mathrm{Embryo}$. (a) $N_\mathrm{Embryo} = 10$. (b) $N_\mathrm{Embryo} = 100$.}
     \label{fig:occurence_rates}
\end{figure}

The distribution of planets changes with the assumption of the initial number of embryos. While Earth-like planets are the most common type for the $N_\mathrm{Embryo} = 100$ simulation, there are more super-Earths than Earths in the $N_\mathrm{Embryo} = 10$ case. As the number of initial embryos decreases, the number of sub-giant and giant planets in the halo increases, since there is proportionally more material available for them to form and grow. We show the occurrence rate, as a function of planet type and stellar population, for the $N_\mathrm{Embryo} = 10$ and $N_\mathrm{Embryo} = 100$ case in Fig. \ref{fig:occurence_rates}.

We also investigate the effect of a metallicity limit for planet formation. \citet{Andama2024} suggest planetesimal formation is severely inhibited for [Fe/H] < -0.6. As an extreme case, we remove all planets around these low metallicity stars, which has a strong influence on the planet population in the halo. For $N_\mathrm{Embryo} = 10$, the number of super-Earths decreases by a factor of eleven. The same holds true for Earth-like planets in the $N_\mathrm{Embryo} = 100$ case. The number of giant planets is less affected by this limit, with a decrease from 34 to 14 in the $N_\mathrm{Embryo} = 10$ case, and no changes in the $N_\mathrm{Embryo} = 100$. This is caused by the fact that massive planets are inherently rare around these low-metallicity stars in the NGPPS population, independent of a metallicity threshold.

\begin{table*}[t]
    \centering
    \caption{Median number of planets for the targets in the LOPS2 and LOPN1 fields across 300 mock simulations, by the number of embryos assumed in the NGPPS populations.}
    \begin{tabularx}{\textwidth}{ll|R|R|R|R}
     &  & $N_\mathrm{Embryo} =10$  & $N_\mathrm{Embryo} =20$ & $N_\mathrm{Embryo} =50$ & $N_\mathrm{Embryo} =100$ \\
     \addlinespace
     \hline \hline
     \addlinespace
     \multirow{5}{*}{Thin Disk} & Earth & 193,446 & 342,480 & 666,603 & 760,515 \\
     & Super-Earth & 212,649 & 313,580 & 600,428 & 732,959 \\
     & Neptunian & 61,418 & 64,345 & 53,561 & 64,353 \\
     & Sub-Giant & 18,955 & 19,474 & 16,883 & 15,740 \\
     & Giant & 48,583 & 50,645 & 37,548 & 42,120 \\
     \addlinespace
     \hline
     \addlinespace
     \multirow{5}{*}{Thick Disk} & Earth & 9,277 & 17,845 & 36,489 & 45,103 \\
     & Super-Earth & 14,410 & 18,260 & 19,868 & 23,812 \\
     & Neptunian & 3,044 & 3,319 & 1,326 & 1,420 \\
     & Sub-Giant & 1,504 & 1,245 & 324 & 322 \\
     & Giant & 1,998 & 1,947 & 722 & 887 \\
     \addlinespace
     \hline
     \addlinespace
     \multirow{5}{*}{Halo (Without [Fe/H] Cut)} & Earth & $425^{+20}_{-19}$ & $887^{+37}_{-43}$ & $2,004^{+56}_{-67}$ & $2,932^{+80}_{-68}$ \\
     & Super-Earth & $1,330^{+28}_{-31}$ & $1,311^{+35}_{-36}$ & $387^{+32}_{-28}$ & $426^{+36}_{-34}$ \\
     & Neptunian & $86^{+9}_{-9}$ & $187^{+13}_{-12}$ & $3^{+2}_{-1}$ & $3^{+3}_{-2}$ \\
     & Sub-Giant & $169^{+10}_{-9}$ & $54^{+10}_{-8}$ & $1^{+1}_{-0}$ & $1^{+1}_{-0}$ \\
     & Giant & $34^{+8}_{-6}$ & $32^{+8}_{-7}$ & $1^{+1}_{-0}$ & $2^{+1}_{-0}$ \\
    \addlinespace
    \hline
    \addlinespace
    \multirow{5}{*}{Halo (With [Fe/H] Cut)} & Earth & $69^{+9}_{-9}$ & $137^{+19}_{-17}$ & $278^{+30}_{-27}$ & $364^{+28}_{-28}$ \\
     & Super-Earth & $120^{+13}_{-11}$ & $143^{+14}_{-13}$ & $104^{+17}_{-14}$ & $125^{+17}_{-21}$ \\
     & Neptunian & $22^{+6}_{-5}$ & $23^{+5}_{-5}$ & $3^{+1}_{-2}$ & $3^{+2}_{-2}$ \\
     & Sub-Giant & $12^{+3}_{-3}$ & $11^{+4}_{-4}$ & $1^{+1}_{-0}$ & $1^{+0}_{-0}$ \\
     & Giant & $14^{+5}_{-5}$ & $13^{+4}_{-4}$ & $1^{+1}_{-0}$ & $2^{+1}_{-0}$ \\
    \end{tabularx}
    \label{tab:number_of_planets}
    \tablefoot{For the halo stars, the 16th-84th percentile are given. The thick and thin disk components have a larger number of stars, resulting only in small variation between mocks ($< 1$\%). For the last, a lower metallicity limit at [Fe/H] < -0.6 is assumed, and planets around lower metallicity stars are removed.}
\end{table*}

\subsection{Expected PLATO planet yields}
To estimate the yield of planet detections for PLATO, we combine the planet populations from Sect. \ref{subsec:planet_populations} with the detection efficiencies from Sect. \ref{subsec:detection_eff_results}. We consider various scenarios, varying the number of initial embryos in the planet formation model (10, 20, 50, and 100) and whether we impose a minimum metallicity threshold of [Fe/H] = -0.6, below which planet formation is assumed to be suppressed.

The resulting planet yields, expressed as the mean number of expected detections for different planet types and stellar populations, are presented in Fig. \ref{fig:observation_heatmaps}. These estimates are for a 2+2 year observation strategy across the LOPS2 and LOPN1 fields. The heat maps show the expected number of detections as a function of orbital period and planet radius.

In general, we find that the highest planet yields are for super-Earths around thin disk stars, with the number of detections decreasing for larger planets, and for planets around thick disk and halo stars. The most populated bin for thin and thick disks are planets with $R = 2$ - 10 $R_\Earth$, and orbital periods between 2 and 5 days. For halo stars, detections are more likely for planets with a similar radius but longer orbital periods of 20–50 days. The yields are also sensitive to the assumed number of initial embryos. A larger number of embryos leads to more planets, although fewer embryos lead on average to more massive, easily detectable planets. As a result, the $N_\mathrm{Embryo}=10$ and $20$ simulation runs lead to the largest number of expected detections at 14,481 and 16,361 respectively. The $N_\mathrm{Embryo}=100$ case on the other hand leads to 13,857 expected detections (Table \ref{tab:planet_yield}).

Imposing a minimum metallicity threshold of [Fe/H] = -0.6, has a large effect on the detection rate around halo stars, as expected (Table \ref{tab:planet_yield_metallicity_limit}). While up to 84 planets can be expected around halo stars without the limit, this number decreases to less than 10 if the metallicity threshold is imposed. In the $N_\mathrm{Embryo} = 100$ case, 23 out of the 300 mock observations contain zero detection of planets around halo stars, if no metallicity limit is imposed. With the metallicity limit, 141 out of 300 planets yield non-detections. This means that while it is likely that PLATO will detect the first exoplanet around a halo star, it is by no means a given. Moreover, a non-detection would not necessarily provide strong evidence for a lower metallicity limit for planet formation around halo stars.

\begin{table}
    \centering
    \caption{Median number of expected planet detections in the different Galactic components across the LOPS2 and LOPN1 fields, for a varying number of initial embryos $N_\mathrm{Embryo}$.}
    \begin{tabular}{l|r|r|r|r}
        Number of Embryos & 10 & 20 & 50 & 100 \\
        \addlinespace
        \hline\hline
        \addlinespace
        Thin Disk & 13,674 & 15,343 & 15,330 & 13,454 \\
        Thick Disk & 761 & 934 & 444 & 399 \\
        Halo & 46 & 84 & 4 & 4 \\
        \addlinespace  
        Total & 14,481 & 16,361 & 15,778 & 13,857 \\
        $\mathrm{[Fe/H]}~ <-0.6$ & 203 & 380 & 23 & 34 \\
    \end{tabular}
    \label{tab:planet_yield}
\end{table}

\begin{table}
    \centering
    \caption{Median number of expected planet detections across the LOPS2 and LOPN1 fields, enforcing the metallicity limit. Planets around stars with [Fe/H] < -0.6 are removed.}
    \begin{tabular}{l|r|r|r|r}
        Number of Embryos & 10 & 20 & 50 & 100 \\
        \addlinespace
        \hline\hline
        \addlinespace
        Thin Disk & 13,636 & 15,260 & 15,320 & 13,432 \\
        Thick Disk & 637 & 715 & 434 & 390 \\
        Halo & 5 & 6 & 1 & 1 \\
        \addlinespace
        Total & 14,278 & 15,981 & 15,755 & 13,823 \\
    \end{tabular}
    \label{tab:planet_yield_metallicity_limit}
\end{table}

\section{Discussion}
\label{sec:discussion}
\subsection{General findings and comparison to previous work}
\label{subsec:general_discussion}
Our estimates predict the detection of between 13,000 and 17,000 planets for a 2+2 year observation strategy across the LOPS2 and LOPN1 fields. This number only accounts for stars classified as thin disk, thick disk, or halo, representing about 51\% of the targets in the full asPIC catalogue. The vast majority (>90\%) of these detections are expected in the thin disk, with 400–1000 planets predicted for the thick disk and between 1 and 80 planets for the halo.

Most detected planets are likely to be super-Earths and sub-Neptunes ($R=2$ - $10 R_\Earth$) with orbital periods between 1 and 100 days. These planets strike a balance between occurrence rate and detectability. We also anticipate 20-130 Earth-like planets with orbital periods of 250–500 days, some of which will be rocky planets in the habitable zone, PLATO's primary target. Our estimates are compatible with \citet{Heller2022}, who predict 11–34 planets with radii between 0.5 and 1.5 $R_\Earth$ in the habitable zone. Our larger upper limit estimate results from the fact that we use a simpler model for stellar variability and planet detection, as well as a broader mass bin ($M = 0.5$ - $2 M_\Earth$) for Earth-like planets.

Our estimates align with \citet{Matuszewski2023b}, who used a similar methodology based on the NGPPS sample. We find similar numbers for the planet yield, with a slightly lower number for us (when scaled to the full asPIC sample) due to our treatment of the stellar radius distribution and limb-darkening effects. It is worth stressing that final planet yield strongly depends on the assumed underlying planet population. Using the NGPPS populations, we find an overall occurrence rate of 4.6 for planets with $M > 0.5~M_\Earth$ and $P<500$ days in the $N_\mathrm{Embryo}=100$ case, and an occurrence rate of 2.6 in the $N_\mathrm{Embryo}=10$ case. This is in contrast to empirical estimates by \citet{Hsu2019b} and \citet{Kunimoto2020c}. \citet{Hsu2019b} find an upper limit on the occurrence rate of about 5 planets per star, primarily due to large uncertainty on the occurrence rate of small ($0.5 R_\Earth$ < $R$ < $1 R_\Earth$) planets with orbits between 64 and 500 days. Exploring this part of the parameter space is the prime goal of PLATO. \citet{Kunimoto2020c} find a much lower occurrence rate of around one planet per star for orbital periods $<400$ days, since they expect a much lower fraction of small ($< 2 R_\Earth$) planets. While the occurrence rate of super-Earths with radii between 2 and 3 $R_\Earth$ are similar between the different estimates, NGPPS predicts a larger fraction of close-in planets ($P < 10$ days), which results in a large fraction of our predicted planet yield being these types of planets.

While the number of planets formed in the simulation varies with the initial number of embryos, the expected number of planet detections remains relatively consistent. The model predicting the highest number of planets ($N_\mathrm{Embryo} =100$) actually yields the lowest number of expected detections. This is because the same amount of disk mass is distributed among more planets in this model, resulting in smaller, less massive planets that are harder to detect. While the median planet radius falls within the super-Earth range, regardless of initial number of embryos, the median radius for the $N_\mathrm{Embryo}=10$ case ($ \overline{R} = 3.1 R_\Earth$, counting only planets with $M > 0.5 M_\Earth$) is 50\% larger than for $N_\mathrm{Embryo}=100$ ($ \overline{R} = 2.1 R_\Earth$).

\subsection{Exoplanet demographics in the Galactic context}
We find characteristic differences in planet populations across Galactic stellar populations, consistent with our previous simulation-based approach \citep{Boettner2024}. The fraction of low-mass planets (Earth and super-Earth) increases as the average metallicity of the stellar population decreases, matching observational findings that average planet masses increase with host star metallicity \citep{Sousa2019}, and that giant planets are more common around metal-rich stars \citep{Thorngren2016a}.

The metal-rich thin disk exhibits a diverse exoplanet population, with Earth-like planets or super-Earths being the most common, depending on the assumed $N_\mathrm{Embryo}$, and hosted in around 90\% of systems. More massive planets (Neptunians, Sub-giants, giants) are rarer, but still found in about 20\% of systems. Compared to the original NGPPS population, our thin disk population is very similar in terms of fraction of planet-hosting systems, occurrence rate, and multiplicity for Earth-like planets, Neptunians, Sub-giants, and giants. This is expected since the NGPPS sample was created to reflect the exoplanet population in the direct solar neighbourhood.

The fraction of giant-hosting systems drops sharply in more metal-poor regions, from 0.16 in the thin disk to 0.004 in the halo. The Earth-hosting fraction remains largely unchanged, while the occurrence rate and multiplicity of Earth-like planets increase. This agrees with previous studies finding that metal-poor stars host a more homogeneous, lower-mass planet population \citep{Petigura2018a}.

The number of giant planets in metal-poor populations depends strongly on the assumed initial number of embryos. Fewer embryos lead to more giant planets due to more available material for single planets, while more embryos lead to more Earth-like planets and fewer giants. For the $N_\mathrm{Embryo}=100$ case, the occurrence rate of Earth-like planets increases with decreasing [Fe/H], while the occurrence rate of super-Earths decreases sharply. For $N_\mathrm{Embryo}=10$ case, the opposite is true (see Fig. \ref{fig:occurence_rates}). In fact, while Earth-like planets are most common in the thin disk, thick disk, and halo for the $N_\mathrm{Embryo}=100$ case, super-Earths are the most common in the $N_\mathrm{Embryo}=10$ case. For $N_\mathrm{Embryo}=20$, the most common type switches, with Earth-like planets being most common in the thin disk and super-Earths being the most common in the thick disk and halo.

\subsection{The impact of host star chemistry on exoplanet populations}
The effect of a star's chemical composition, particularly the abundance of $\alpha$-elements, on planet formation is an active area of research. Theoretical studies suggest that the abundance of water ice in planetary building blocks, and potentially in planets themselves, depends on both the metallicity and alpha-element content of the host star \citep{Bitsch2020, Cabral2023}. Observationally, there are indications that the relative abundance of super-Earths and sub-Neptunes may vary with $\alpha$-element abundance \citep{Chen2022}, and that giant planets might be more frequent around $\alpha$-enriched, metal-poor stars \citep{Haywood2008, Haywood2009}, compared to $\alpha$-poor, metal-poor stars. However, current observational evidence is limited.

PLATO is poised to significantly expand our understanding in this area by providing a large sample of planets around stars with diverse chemical compositions. We expect PLATO to detect between 400 and 1000 planets around thick disk stars, with an average metallicity of -0.3 to -0.4 and an $\alpha$-elemental abundance around 0.17. This would be a major increase compared to the eight planets around seven thick disk stars found by \citet{Bashi2022} for the \textit{Kepler} sample. The vast majority of these planets are predicted to be short-period (2-20 days) super-Earths and sub-Neptunes ($R = 2$ - $10 R_\Earth$), though with the caveat that is type of planet is over-represented in the NGPPS population compared to other estimates (see Sect. \ref{subsec:general_discussion}). \citet{Bashi2022} find for their limited sample of planets with $R=1$ - $2 R_\Earth$ and $P=1$ - 100 days have an occurrence rate around thin disk stars that is more than twice the occurrence rate around thick disk stars. We find a much smaller deviation of only around 10\% in the $N_\mathrm{Embryo}=100$ case, and basically no deviation in the $N_\mathrm{Embryo}=10$ case.
The increased number statistics that PLATO provides will help establish if the findings by \citet{Bashi2022} are a statistical fluke, or if the dynamical history of the star has a direct effect on the occurrence rate as they have suggested.

Within the NGPPS population there is only a small hint of a radius valley \citep{Fulton2017} between super-Earths ($R$ between 1–1.7 $R_\Earth$) and sub-Neptunes ($R$ between 2.1 and 3.5 $R_\Earth$) with periods < 100 days (see also \citealt{Burn2021}, with improvements by \citealt{Burn2024}). This valley is largely unchanging with metallicity, and between thin disk and thick disk, in contrast to the metallicity evolution suggested by \citet{Chen2022}. They find that the valley disappears for lower metallicity hosts, as the number of super-Earths decreases. For fixed [Fe/H], they further find that the ratio of super-Earths and sub-Neptunes decreases as the $\alpha$-elemental abundance increases. The regime probed by PLATO is therefore ideal for studying the evolution of the planet radius valley, and analysing the effects of chemical abundances on the formation of rocky versus gaseous and water-rich versus dry worlds.

We expect to find a smaller number of up to 100 of Neptune and sub-giant-sized planets in the thick disk, mostly with periods < 50 days. Additionally, we anticipate detecting up to a dozen giant planets around thick disk stars, although roughly half of these are predicted to have longer orbital periods (> 100 days). However, it is worth noting that the fraction of high metallicity stars ([Fe/H] > 0) is elevated in our thick disk sample, due to thin disk contamination, which is likely to artificially increase the occurrence rate of large planets in this sample. Nonetheless, comparing the occurrence rates of large planets around low-metallicity stars will help us understand if high $\alpha$-content aids planet formation, as suggested by \citet{Gundlach2015}.

\subsection{Planets around stars with sub-solar metallicity}
The formation of planets around low-metallicity stars is a topic of ongoing research. Theoretical and observational evidence suggests a potential minimum metallicity threshold for planet formation, with estimates ranging from [Fe/H] = -0.5 \citep{Mortier2012, Mordasini2012} to -1.5 \citep{Johnson2012a}. To date, only a handful of planets around low-metallicity stars have been discovered. A recent study by \citet{Andama2024} suggests that planetesimal formation, is significantly hindered in low-metallicity environments, particularly for [Fe/H] < -0.6.

PLATO has the potential to greatly improve our understanding of this regime of planet formation. A lower metallicity limit, as suggested by \citet{Andama2024}, would greatly affect planet yield in low-metallicity environments. Across the LOPS2 and LOPN1 fields, we were able to classify 3,451 target stars with [Fe/H] < -0.6 into the thin disk, thick disk or halo. The expected number of planet detections around these targets varies widely. From zero in the presence of a metallicity limit, to around 30 in the $N_\mathrm{Embryo}=100$ case. This value can reach up to 400 in the $N_\mathrm{Embryo}=20$ case, which has on average more massive, more easily detectable planets, especially in low-metallicity disks where the available solid material is limited. The detection (or non-detection) of planets around these stars can yield valuable constraints for planet formation models in the low-metallicity regime. 

 We predict between 1 and 80 planet detections around halo stars, with metallicities < -0.3. Most of these planets are expected to be super-Earths with periods between 2 and 50 days. We want to especially highlight a special target list of 47 halo targets in the high-priority, high-S/N PLATO P1 sample, which are prime targets in the search for planets around halo stars. The sample has an estimated median metallicity of -1.27, although metallicity estimates less reliable for values this low. The highest metallicity star in this sample has [Fe/H] = -0.324. A single star in this sample (Gaia DR3 ID: 4676601464106231040) has an APOGEE-confirmed metallicity estimate of [Fe/H] = -1.467. The full target list can be found in Table \ref{tab:LOPS2_special_targets} and Table \ref{tab:LOPN1_special_targets}.

\section{Conclusion}
\label{sec:conclusions}
In this study, we have demonstrated the potential of the upcoming PLATO mission to greatly improve our understanding of exoplanet demographics across diverse Galactic stellar populations. By combining asPIC with a probabilistic kinematic classification scheme and the NGPPS planet population synthesis models, we have estimated the PLATO exoplanet yields for the thin disk, thick disk, and stellar halo across the LOPS2 and LOPN1 fields for a 2+2 year observation strategy. Our primary findings are:
\begin{enumerate}
    \item PLATO is expected to detect a vast number of exoplanets, primarily super-Earths and sub-Neptunes, orbiting stars within the thin disk, thick disk, and halo. We predict the detection of at least 13,000 planets across these populations, with the majority residing in the thin disk. This extensive dataset will enable detailed studies of exoplanet occurrence rates, architectures, and properties as a function of Galactic location and stellar population.
    \item The exoplanet populations exhibit distinct characteristics across different Galactic components. The thin disk is expected to host a diverse range of planets, including a significant number of habitable-zone rocky planets. The thick disk, with its lower average metallicity, is predicted to have a higher relative fraction of low-mass planets (Earth-like and super-Earths) and a lower occurrence rate of giant planets.
    \item PLATO will be able to observe at least 400 planets around stars in the $\alpha$-enriched thick disk. These observations will enable detailed investigations of the relationship between $\alpha$-element abundance, the properties of planet building blocks, and planet populations. This will provide insights into the role of $\alpha$ elements in planet formation and the potential diversity of planet compositions across different Galactic environments.
    \item We predict the detection of several dozen planets around halo stars, with [Fe/H] < - 0.3, and potentially hundreds with [Fe/H] < -0.6 across the thin disk, thick disk, and halo. These observations will help  constrain the possible existence of a minimum metallicity threshold for planet formation.
    \item We have identified a list of 47 targets that have been kinematically classified as halo stars in the high-priority, high-S/N PLATO P1 sample across the LOPS2 and LOPN1 fields, making them prime candidates in the search for planets around low-metallicity stars.
\end{enumerate}
These results demonstrate the potential of PLATO to significantly advance our understanding of planet formation and evolution across different stellar populations in the Milky Way. The list of halo target stars identified by us can be found in Table \ref{tab:LOPS2_special_targets} and Table \ref{tab:LOPN1_special_targets}.

\section*{Data Availability}
The full catalogue of stars, classified by their Galactic component membership, a target list of the LOPS2 and LOPN1 fields, as well as the halo special target list, can be found at \url{https://doi.org/10.5281/zenodo.11428968}. The analysis code, including notebooks to recreate all figures in this work, are located at \url{https://github.com/ChrisBoettner/plato}.

\begin{acknowledgements}
CB thanks the Young Academy Groningen for their generous support through an interdisciplinary PhD fellowship. PD acknowledges support from the NWO grant 016.VIDI.189.162 (``ODIN") and warmly thanks the European Commission's and University of Groningen's CO-FUND Rosalind Franklin program. 

We want to thank Alexandre Emsenhuber for providing us with the Monte Carlo variable dataset used for the original NGPPS sample, and the Data \& Analysis Center for Exoplanets (\url{dace.unige.ch}) for seamless access to the NGPPS and observational exoplanet data. Valerio Nascimbeni and Hugh Osborn have helped with access to the platopoint code used in this analysis. We further want to thank Else Starkenburg for her helpful comments.

This work has made use of data from the European Space Agency (ESA) mission
{\it Gaia} (\url{https://www.cosmos.esa.int/gaia}), processed by the {\it Gaia}
Data Processing and Analysis Consortium (DPAC,
\url{https://www.cosmos.esa.int/web/gaia/dpac/consortium}). Funding for the DPAC
has been provided by national institutions, in particular the institutions
participating in the {\it Gaia} Multilateral Agreement.

This work made use of the Third Data Release of the GALAH Survey (Buder et al. 2021). The GALAH Survey is based on data acquired through the Australian Astronomical Observatory, under programs: A/2013B/13 (The GALAH pilot survey); A/2014A/25, A/2015A/19, A2017A/18 (The GALAH survey phase 1); A2018A/18 (Open clusters with HERMES); A2019A/1 (Hierarchical star formation in Ori OB1); A2019A/15 (The GALAH survey phase 2); A/2015B/19, A/2016A/22, A/2016B/10, A/2017B/16, A/2018B/15 (The HERMES-TESS program); and A/2015A/3, A/2015B/1, A/2015B/19, A/2016A/22, A/2016B/12, A/2017A/14 (The HERMES K2-follow-up program). We acknowledge the traditional owners of the land on which the AAT stands, the Gamilaraay people, and pay our respects to elders past and present. This paper includes data that has been provided by AAO Data Central (datacentral.org.au).

Funding for the Sloan Digital Sky 
Survey IV has been provided by the 
Alfred P. Sloan Foundation, the U.S. 
Department of Energy Office of 
Science, and the Participating 
Institutions. 

SDSS-IV acknowledges support and 
resources from the Center for High 
Performance Computing  at the 
University of Utah. The SDSS 
website is www.sdss4.org.

SDSS-IV is managed by the 
Astrophysical Research Consortium 
for the Participating Institutions 
of the SDSS Collaboration including 
the Brazilian Participation Group, 
the Carnegie Institution for Science, 
Carnegie Mellon University, Center for 
Astrophysics | Harvard \& 
Smithsonian, the Chilean Participation 
Group, the French Participation Group, 
Instituto de Astrof\'isica de 
Canarias, The Johns Hopkins 
University, Kavli Institute for the 
Physics and Mathematics of the 
Universe (IPMU) / University of 
Tokyo, the Korean Participation Group, 
Lawrence Berkeley National Laboratory, 
Leibniz Institut f\"ur Astrophysik 
Potsdam (AIP),  Max-Planck-Institut 
f\"ur Astronomie (MPIA Heidelberg), 
Max-Planck-Institut f\"ur 
Astrophysik (MPA Garching), 
Max-Planck-Institut f\"ur 
Extraterrestrische Physik (MPE), 
National Astronomical Observatories of 
China, New Mexico State University, 
New York University, University of 
Notre Dame, Observat\'ario 
Nacional / MCTI, The Ohio State 
University, Pennsylvania State 
University, Shanghai 
Astronomical Observatory, United 
Kingdom Participation Group, 
Universidad Nacional Aut\'onoma 
de M\'exico, University of Arizona, 
University of Colorado Boulder, 
University of Oxford, University of 
Portsmouth, University of Utah, 
University of Virginia, University 
of Washington, University of 
Wisconsin, Vanderbilt University, 
and Yale University.

The analysis has been performed using Python, and the scientific data analysis libraries \texttt{numpy} \citep{Harris2020}, \texttt{scipy} \citep{Virtanen2020a}, \texttt{scikit-learn} \citep{Pedregosa2011} and \texttt{pandas} \citep{Thepandasdevelopmentteam2023}. Astronomical calculations and data retrieval were performed using \texttt{astropy} \citep{AstropyCollaboration2022}, \texttt{astroquery} \citep{Ginsburg2019} and \texttt{galpy} \citep{Bovy2015}. Visualisation were made using \texttt{matplotlib} \citep{ThomasACaswell2023} and \texttt{seaborn} \citep{Waskom2021}. The PLATO targets were selected using a modified version of \texttt{platopoint} by Hugh P. Osborn. The code used for this analysis can be found at \url{https://github.com/ChrisBoettner/plato}.

\end{acknowledgements}

\bibliographystyle{aa} 
\bibliography{main} 


\begin{appendix}
\onecolumn
\section{Additional figure}
\begin{figure*}[!htb]
     \centering
     \begin{subfigure}{0.43\paperwidth}
        \centering
        \includegraphics[width=\textwidth]{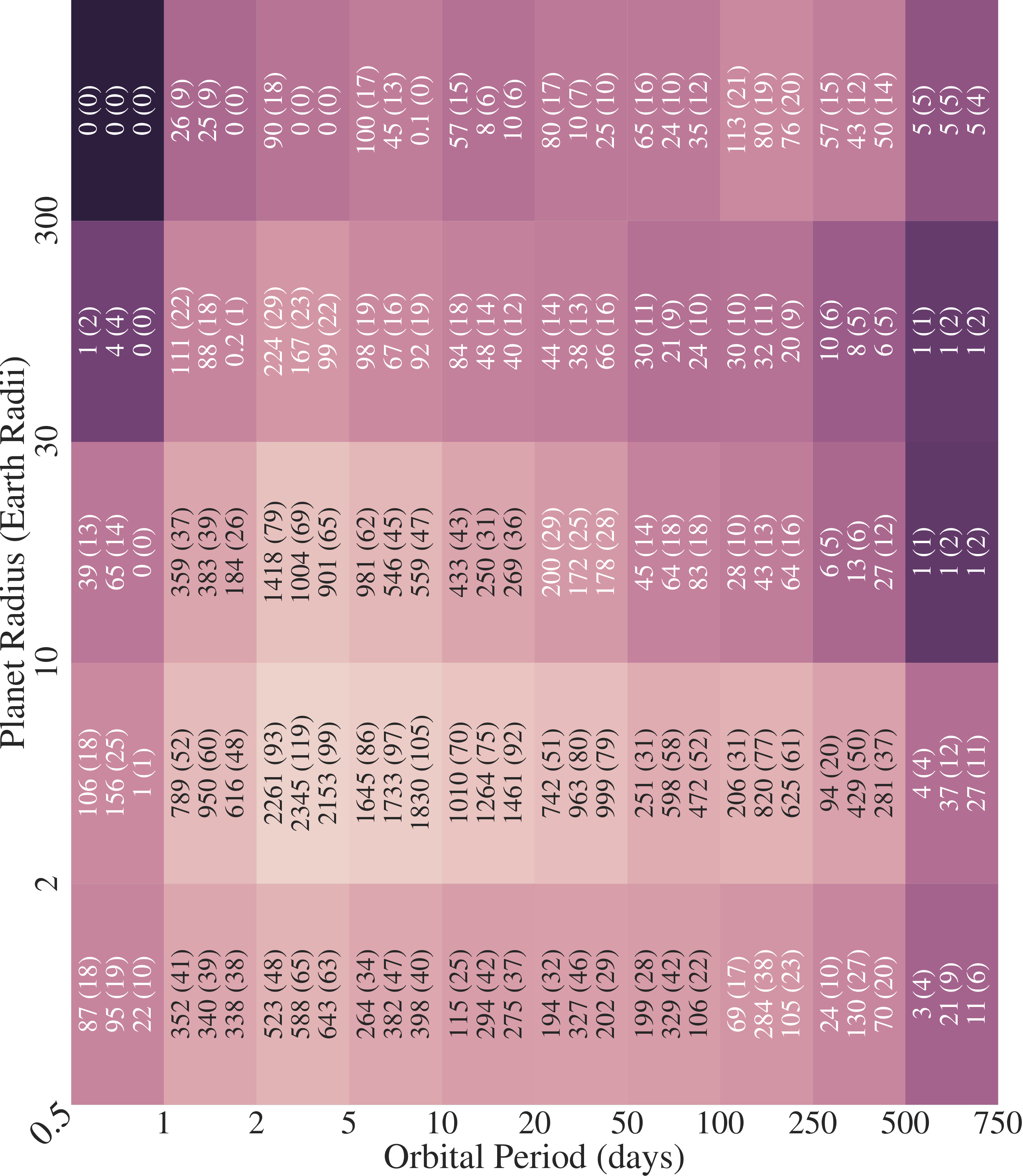}
        \caption{Thin disk.}
        \label{fig:observation_heatmap_thin_disk}
     \end{subfigure}
     \begin{subfigure}{0.43\paperwidth}
        \centering
        \includegraphics[width=\textwidth]{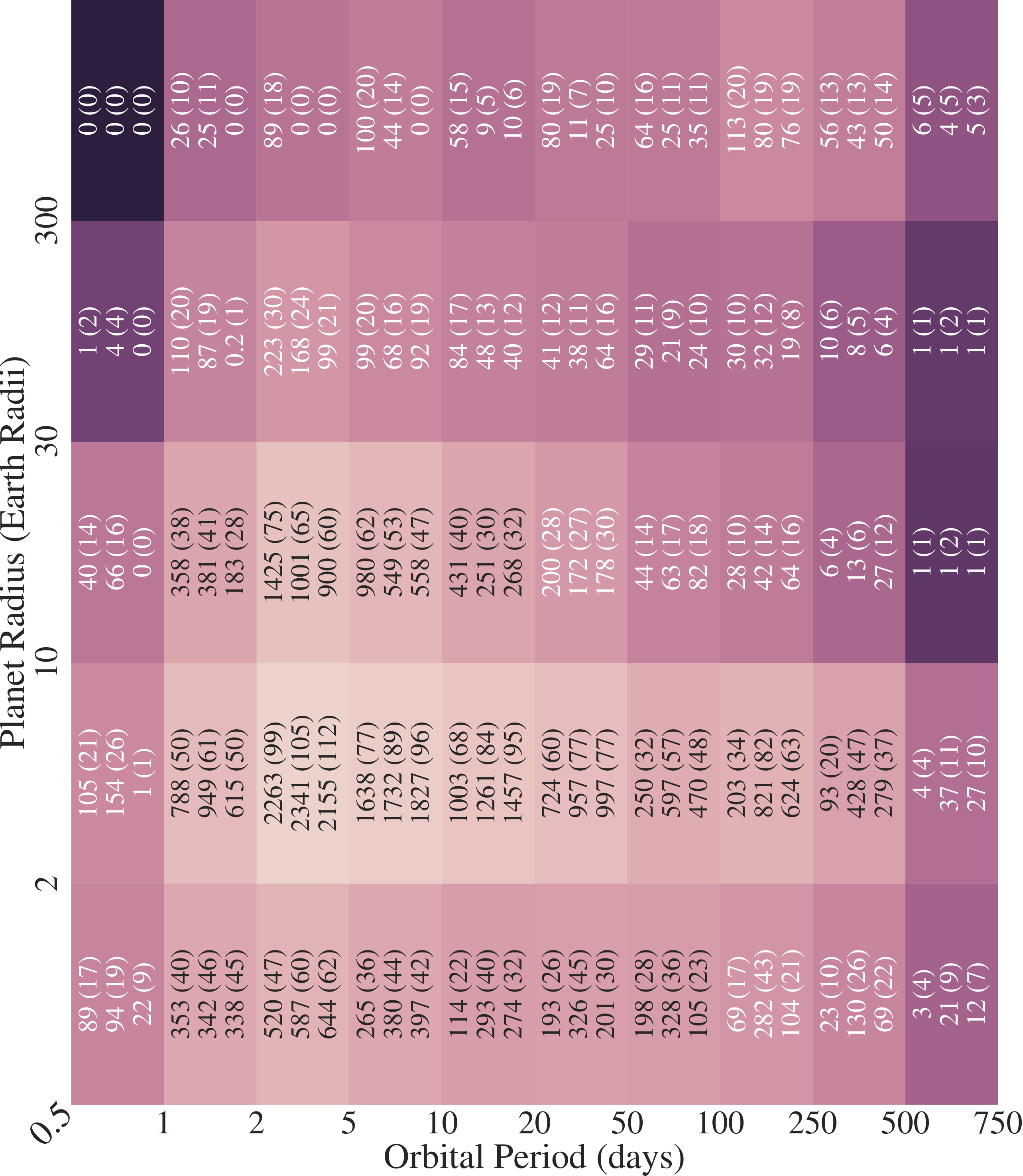}
        \caption{Thin disk (with a [Fe/H] limit).}
        \label{fig:observation_heatmap_thin_disk_metallicity_cut}
     \end{subfigure}\vspace{5mm}
     
    \begin{subfigure}{0.43\paperwidth}
        \centering
        \includegraphics[width=\textwidth]{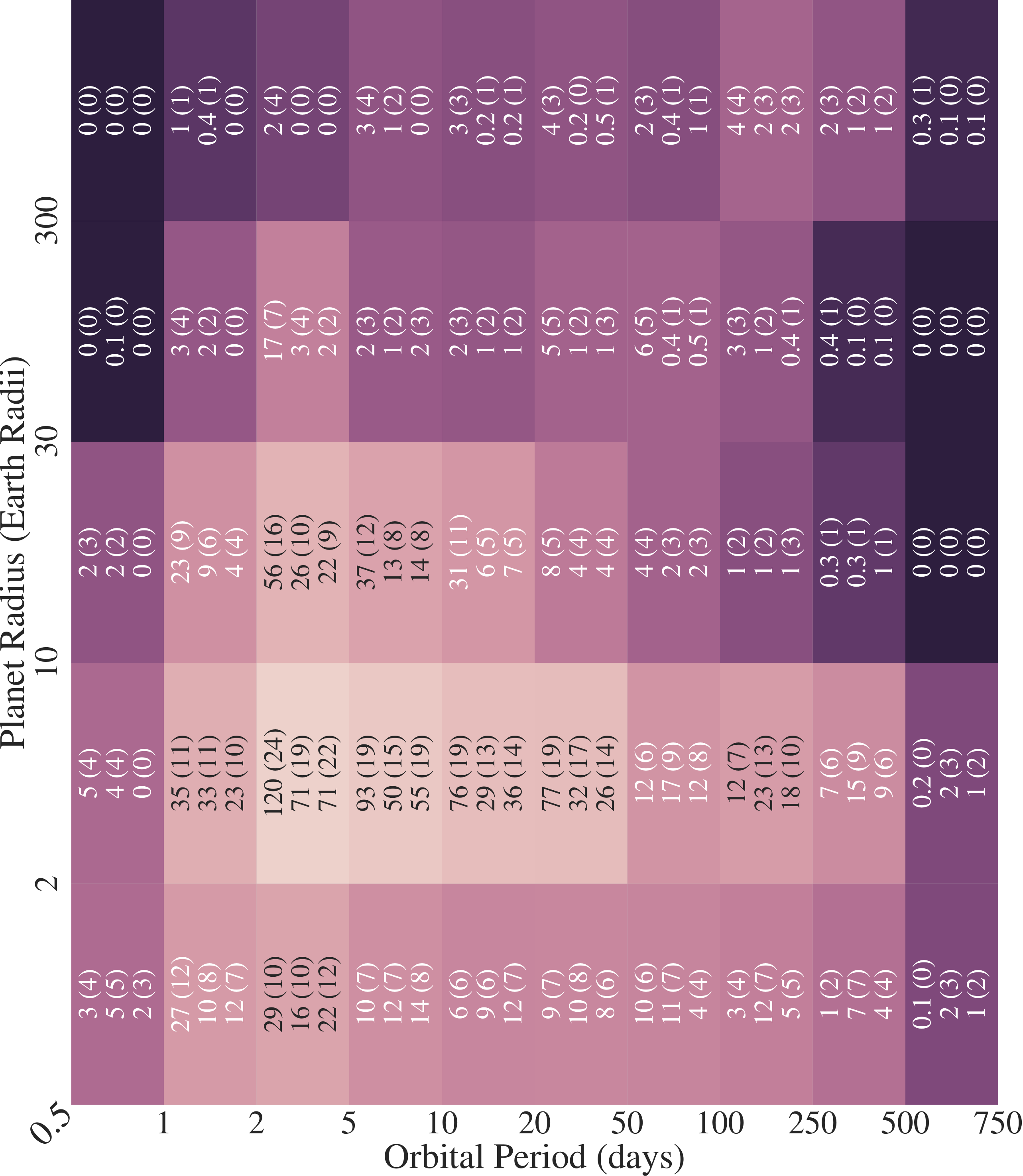}
        \caption{Thick disk.}
        \label{fig:observation_heatmap_thick_disk}
     \end{subfigure}
     \begin{subfigure}{0.43\paperwidth}
        \centering
        \includegraphics[width=\textwidth]{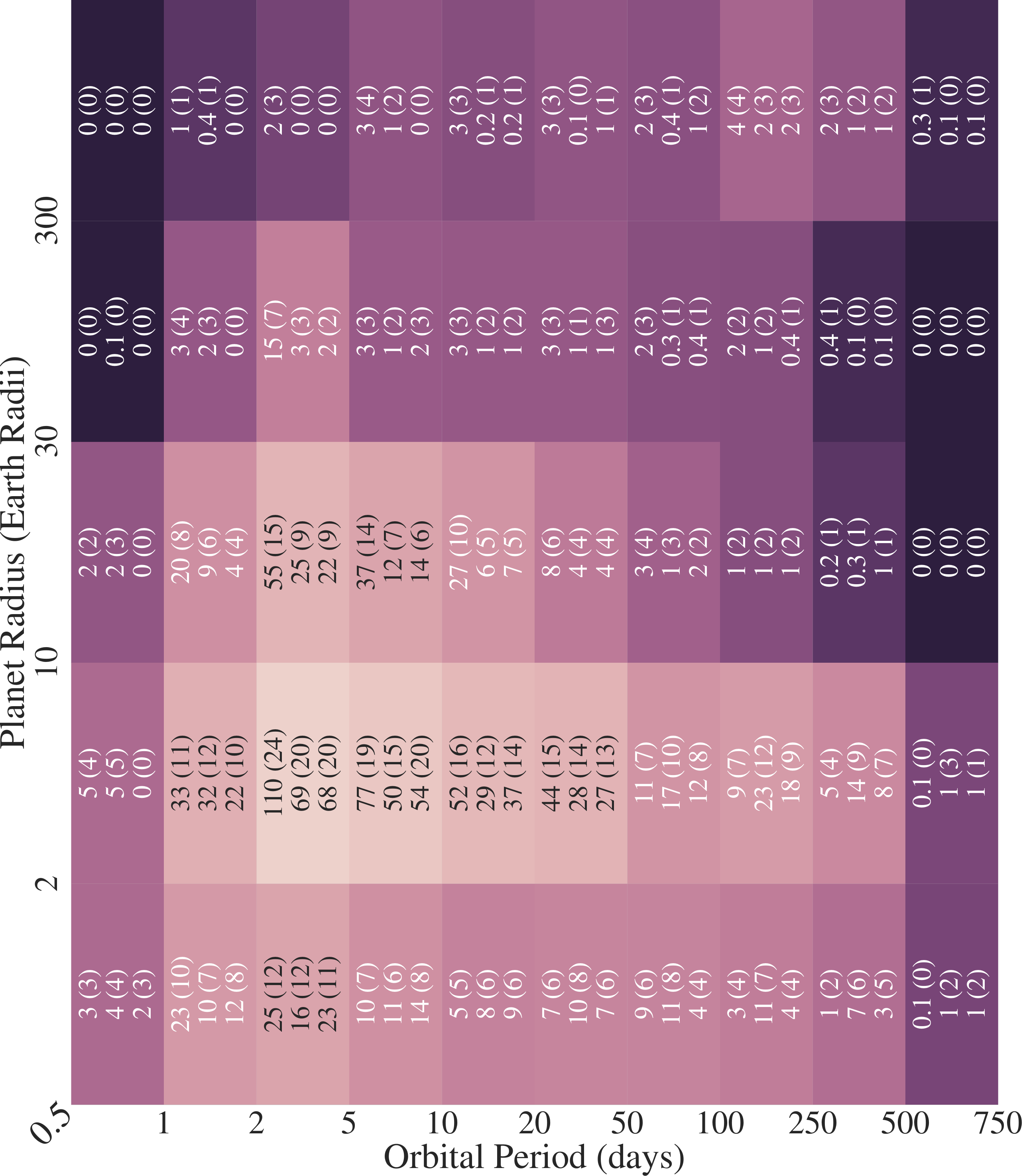}
        \caption{Thick disk (with a [Fe/H] limit).}
        \label{fig:observation_heatmap_thick_disk_metallicity_cut}
     \end{subfigure}
\end{figure*}

\begin{figure*}[!htb]
    \centering
    \ContinuedFloat
    \begin{subfigure}{0.43\paperwidth}
        \centering
        \includegraphics[width=\textwidth]{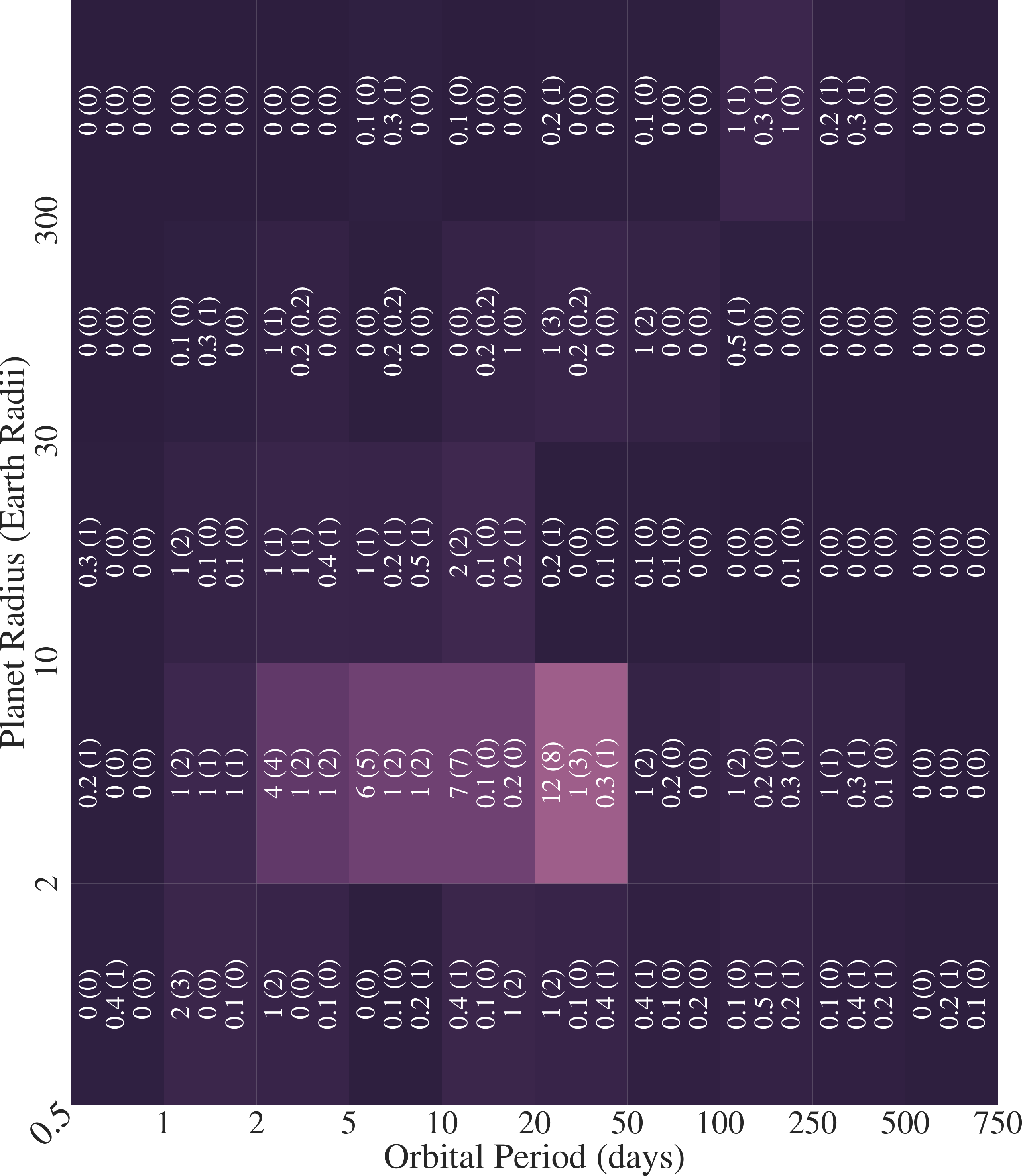}
        \caption{Halo.}
        \label{fig:observation_heatmap_halo}
     \end{subfigure}
     \begin{subfigure}{0.43\paperwidth}
        \centering
        \includegraphics[width=\textwidth]{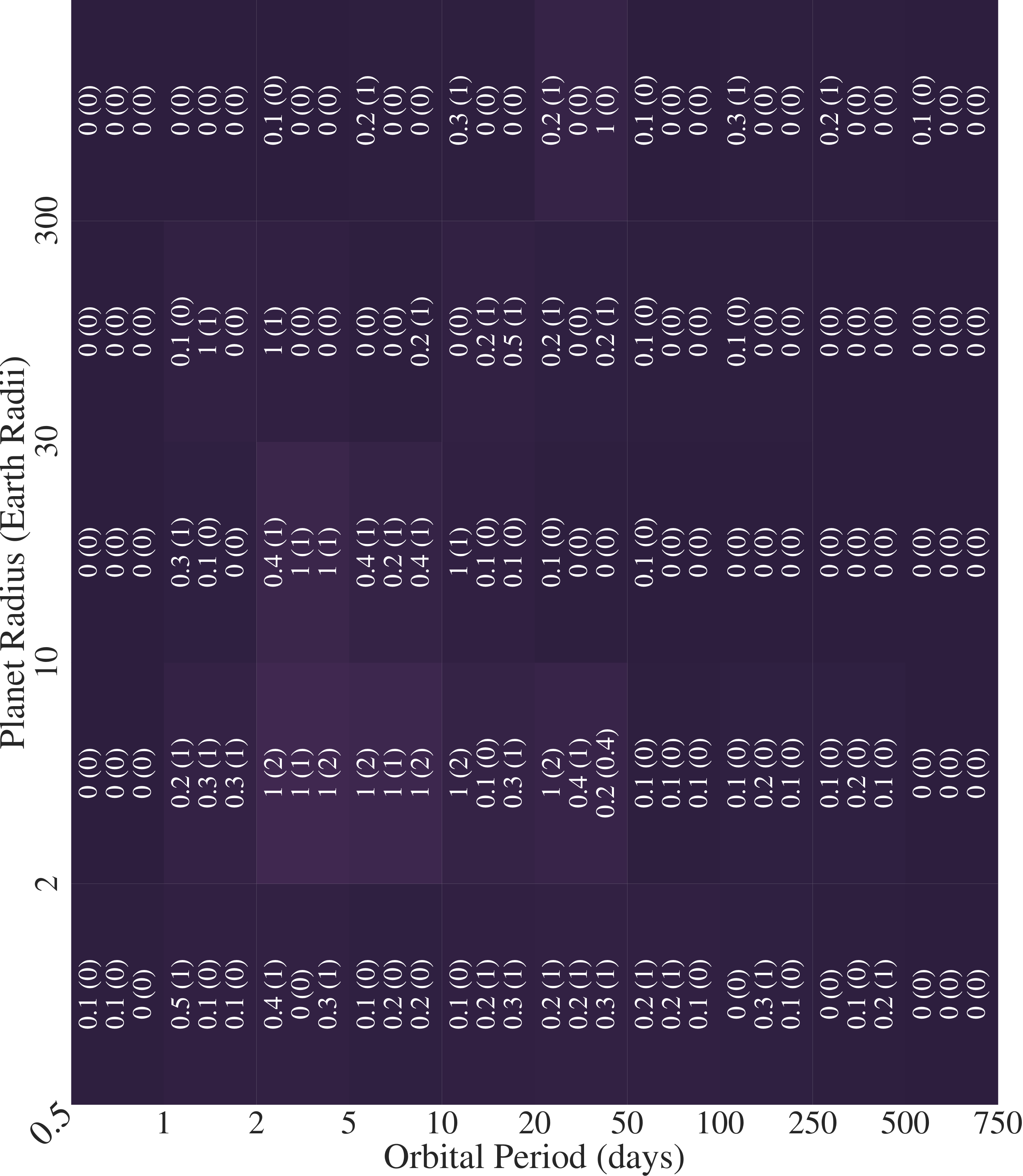}
        \caption{Halo (with a [Fe/H] limit).}
        \label{fig:observation_heatmap_halo_metallicity_cut}
     \end{subfigure}
     \caption{Heat maps of the expected number of observed planets in across the LOPS2 and LOPN1 fields for a 2+2 year observation strategy, as a function of orbital period and planet radius. The numbers in each bin represent the mean and the interquartile range (16th to 84th percentile) for 300 mock observations, for $N_\mathrm{Embryo} = [10, 50, 100]$ (top to bottom). The heat maps on the right side (b, d, f) enforce a minimum stellar metallicity of [Fe/H] = -0.6, below which no planets can form as suggested by \citet{Andama2024}. Note that these estimates are only for the targets we were able to classify into one of the three components (see Table \ref{tab:star_count_per_component}), not considering stars that were categorised as thick disk or halo candidates, or those removed in the sample selection process (Sect. \ref{subsec:stellar_sample}).}
     \label{fig:observation_heatmaps}
\end{figure*}

\twocolumn
\section{Comparison of \textit{Gaia} metallicity and surface gravity with APOGEE and GALAH}
\label{appendixa:gaia_high_res_comparison}
\begin{figure}[ht]
     \centering
     \begin{subfigure}{\columnwidth}
        \centering
        \includegraphics[width=\textwidth]{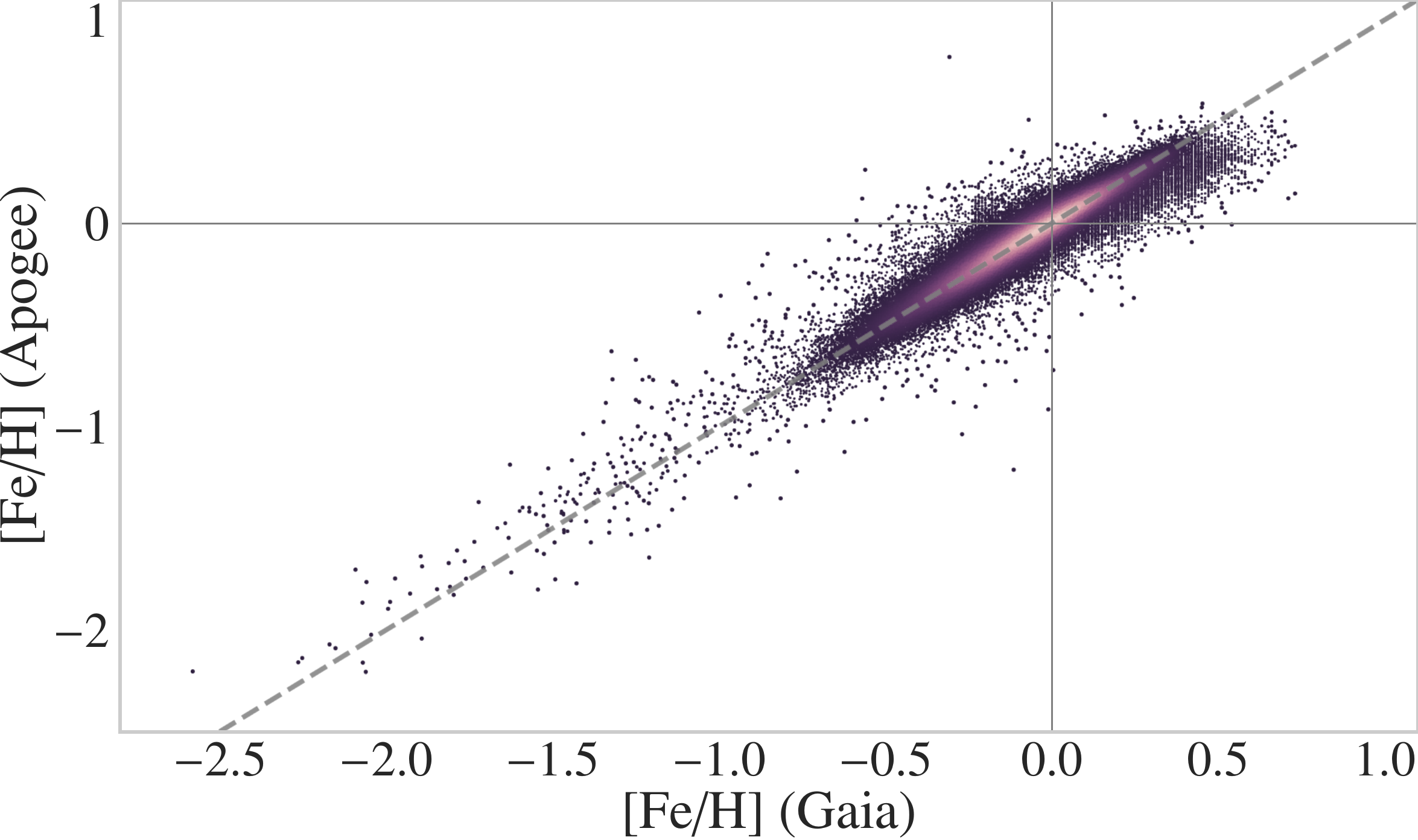}
        \caption{}
        \label{fig:metallicity_scatterplot_apogee}
     \end{subfigure}
     \begin{subfigure}{\columnwidth}
        \centering
        \includegraphics[width=\textwidth]{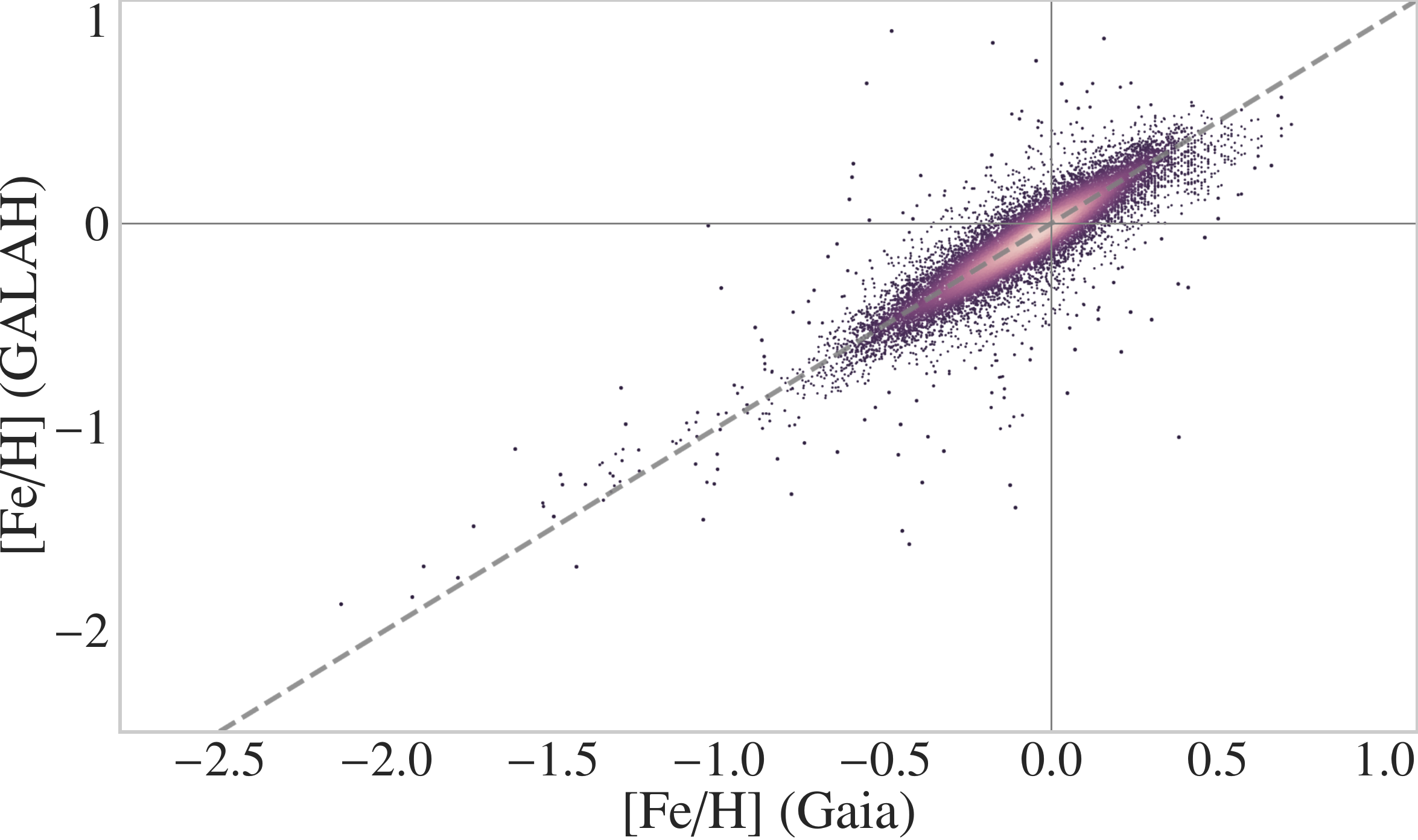}
        \caption{}
        \label{fig:metallicity_scatterplot_galah}
     \end{subfigure}
     \caption{\textit{Gaia} metallicity estimates for our sample compared to values determined from APOGEE and GALAH. Panel (a): APOGEE. Panel (b): GALAH. The \textit{Gaia} estimates show a larger fraction of low- and high-metallicity stars, while the APOGEE and GALAH samples are more centralised around 0. The colour represents the density of points, while the grey dashed line shows the x=y diagonal.}
     \label{fig:metallicity_scatterplot}
\end{figure}

\begin{figure}[ht]
     \centering
     \begin{subfigure}{\columnwidth}
        \centering
        \includegraphics[width=\textwidth]{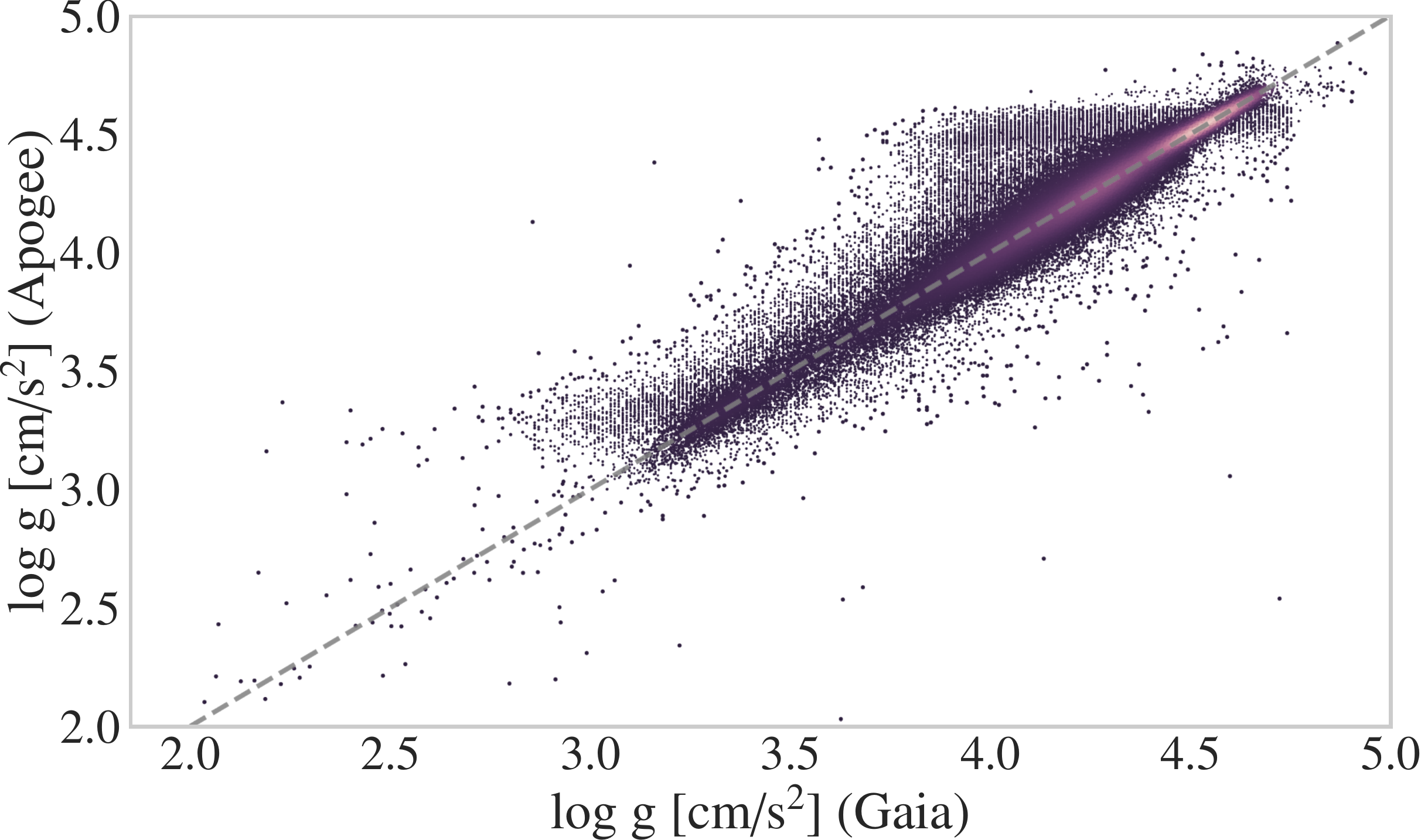}
        \caption{}
        \label{fig:logg_scatterplot_apogee}
     \end{subfigure}
     \begin{subfigure}{\columnwidth}
        \centering
        \includegraphics[width=\textwidth]{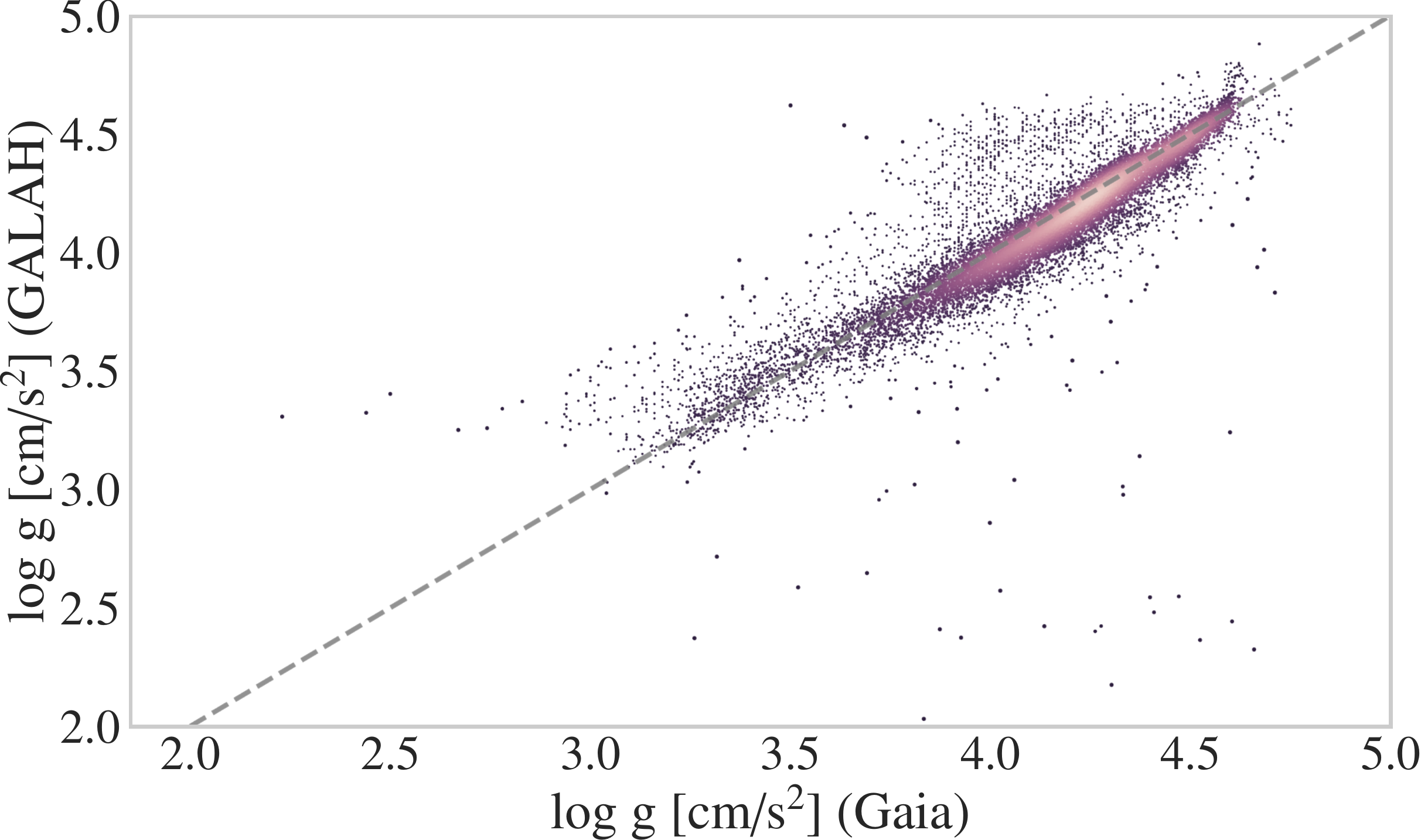}
        \caption{}
        \label{fig:logg_scatterplot_galah}
     \end{subfigure}
     \caption{\textit{Gaia} surface gravity estimates for our sample compared to values determined from APOGEE and GALAH. Panel (a): APOGEE. Panel (b): GALAH. Note the artefacts around 3.7–4.7 cm/s$^2$, where the \textit{Gaia} results have a tendency to underestimate log $g$ for some targets.}
     \label{fig:logg_scatterplot}
\end{figure}

We compare the metallicity estimates derived from \textit{Gaia} (using medium-resolution spectra, and XGBoost, using the quality flags described in Sect. \ref{subsec:stellar_sample}) with those from the APOGEE and GALAH sub-samples, which are based on high-resolution spectra. The \textit{Gaia} estimates show good agreement with the high-resolution values within the metallicity range of -2 to 0.8 (Fig. \ref{fig:metallicity_scatterplot}). The mean metallicity for the \textit{Gaia}, APOGEE and GALAH estimates are -0.096, -0.085 and -0.100, respectively. For surface gravity, the \textit{Gaia} estimates align well with the APOGEE values between 2 and 4.8. For the GALAH sample, the values match between 3 and 4.6. For values below three, the \textit{Gaia} estimates are systematically biased towards larger values (Fig. \ref{fig:logg_scatterplot}).

\section{Effects of stellar variability}
\label{appendixb:stellar_variability}
Variations in stellar flux due to intrinsic effects like granulation are an inherent source of noise in exoplanet transit photometry and spectroscopy. This stellar variability is widely recognised as a major factor limiting the detection capabilities of \textit{Kepler}, which failed to identify a true Earth analogue due to underestimating the activity of FGK stars compared to the Sun \citep{Jenkins2002, Gilliland2011}. The heightened stellar activity in these stars resulted in a photometric precision of 30 ppm over 6.5 hours for \textit{Kepler}, falling short of the 20 ppm goal for a $K_p=12$ star \citep{Jenkins2011}.

While stellar variability poses a significant challenge, studies suggest it should not prevent PLATO from detecting Earth-like planets \citep{Hippke2015, Morris2020}. The extent to which a transit is affected by stellar variability depends on various factors, including the properties of the star and planet, as well as the de-trending method employed. \citet{Morris2020} conducted a comprehensive analysis using the PLATO Solar-like Light-curve Simulator (PSLS, \citealt{Samadi2019}) to assess this impact.

\begin{figure}
    \centering
    \includegraphics[width=\columnwidth]{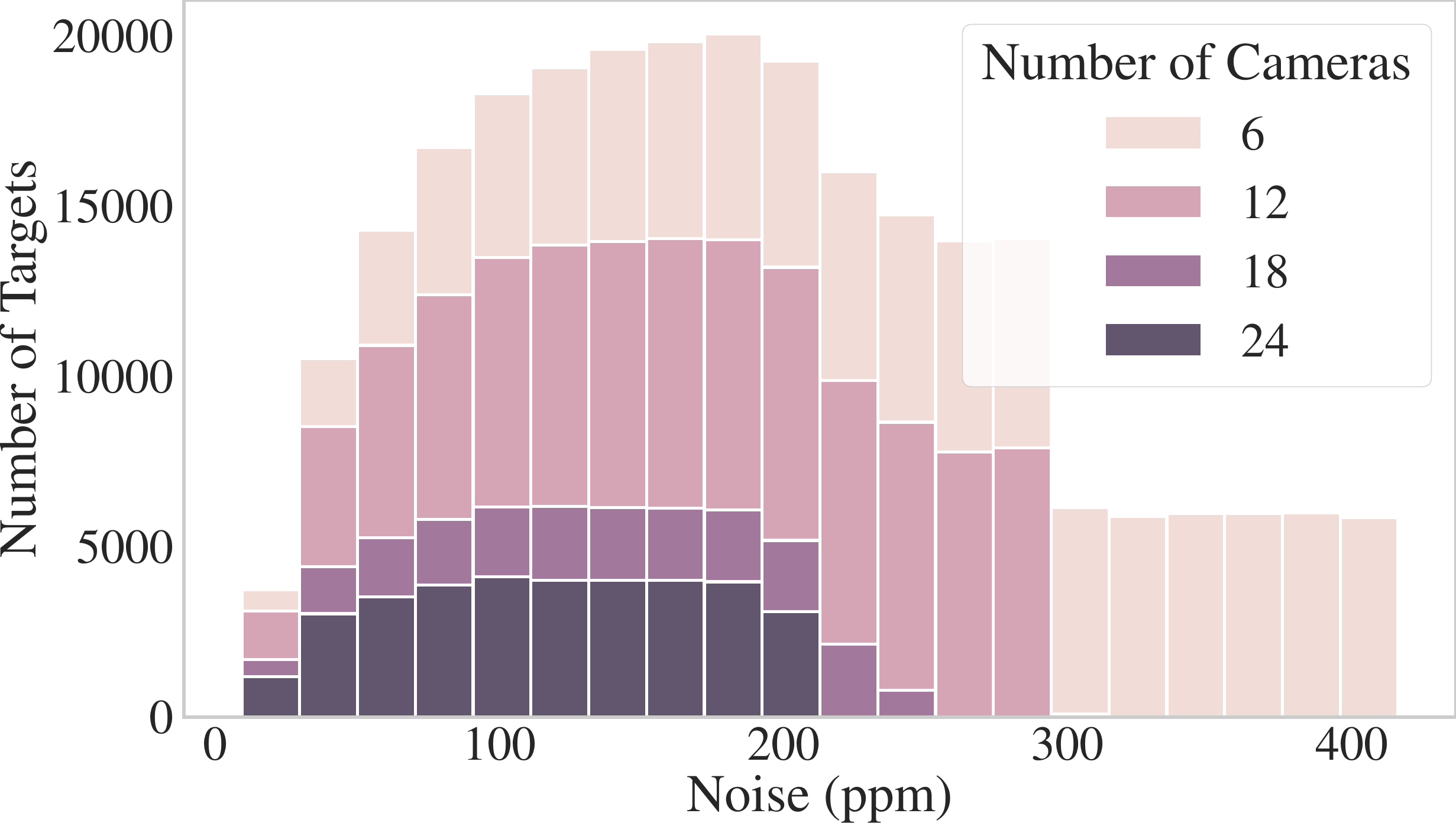}
    \caption{Distribution of estimated total noise (systematic + random + photon), excluding stellar variability, for the target stars in the LOPS2 and LOPN1 fields. The noise modelling is described in Sect. \ref{subsec:noise_model}.}
    \label{fig:noise_hist}
\end{figure}

For this study, we adopted a simplified approach, assuming a stellar variability of 10 ppm for a light curve integrated over one hour. In comparison to systematic, random, and photon noise, this contribution is relatively minor for most targets in the LOPS2 and LOPN1 fields, where the overall noise ranges from 30 ppm to 400 ppm (Fig. \ref{fig:noise_hist}). Stellar variability primarily becomes a concern for detecting small planets at large orbital distances.

\begin{figure}
    \centering
    \includegraphics[width=\columnwidth]{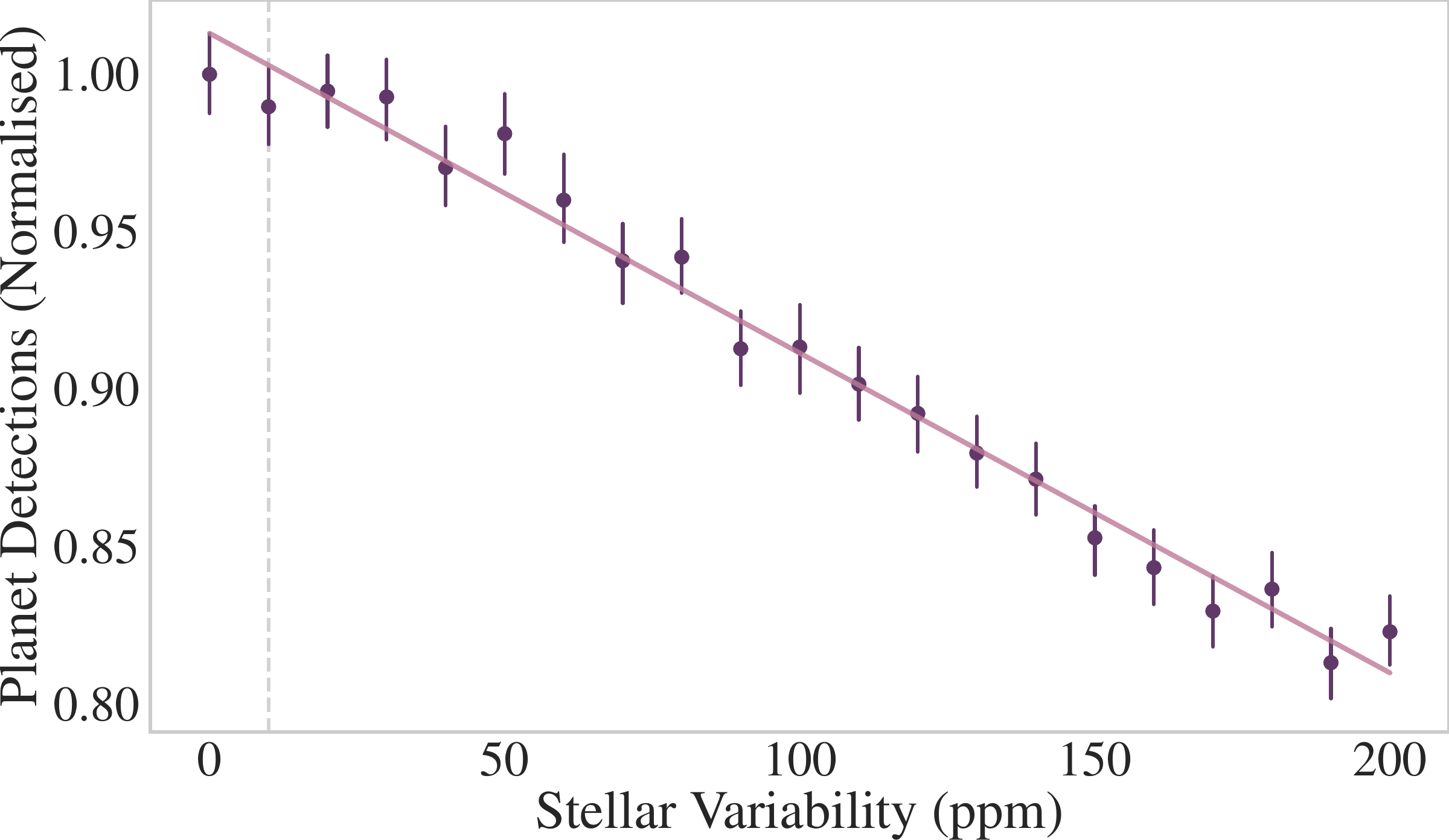}
    \caption{Effects of stellar variability on the number of exoplanet detections. The values are normalised to the number of detections without a stellar variability component ($\sigma_\mathrm{star}=0$). The grey vertical line corresponds to the 10ppm we assume for this work.}
    \label{fig:stellar_variability_plot}
\end{figure}

Figure \ref{fig:stellar_variability_plot} illustrates how the number of detected planets is influenced by stellar variability. Our simple model exhibits a linear decrease in the number of detected planets with increasing stellar variability between 0 and 200 ppm, with a slope of 0.001 ppm$^{-1}$. Even at a variability of 200 ppm, approximately 80\% of planets are still detectable. The missed planets are primarily Earth-like planets with orbital periods exceeding 100 days.

\section{Halo special target list}
\label{appendixc:halo_target_list}
Tables \ref{tab:LOPS2_special_targets} and \ref{tab:LOPN1_special_targets} contain the targets that have been kinematically classified as halo stars, and fall within the high-priority PLATO P1 sample.
 
\begin{table*}
    \centering
    \caption{Special target list for halo stars in the LOPS2 field.}
    \begin{tabularx}{\textwidth}{r|RR|RR|R}
    \textit{Gaia} DR3 ID & RA (hms) & Dec (dms) & Gal. longitude (deg) & Gal. latitude (deg) & V Magnitude\\ 
    \addlinespace
    \hline\hline
    \addlinespace
    2892879693284897536 & 06h16m37.01s & -32d27m15.29s & 239.62354 & -20.99928 & 10.89 \\ \addlinespace
    2911140863435161856 & 05h59m33.51s & -26d16m35.09s & 232.0118 & -22.39392 & 10.91 \\ \addlinespace
    4676601464106231040 & 04h23m43.36s & -62d17m05.00s & 273.55676 & -40.71957 & 10.52 \\ \addlinespace
    4768015406298936960 & 05h47m36.18s & -54d09m57.56s & 261.95891 & -30.85913 & 10.57 \\ \addlinespace
    4772221523607359232 & 05h25m26.69s & -51d58m42.48s & 259.15855 & -34.06364 & 10.98 \\ \addlinespace
    4797516475799429888 & 05h18m19.97s & -48d52m16.70s & 255.33827 & -35.08485 & 10.63 \\ \addlinespace
    4818949324679566720 & 05h13m48.21s & -40d54m03.53s & 245.53908 & -35.2076 & 10.73 \\ \addlinespace
    4864851495597048448 & 04h33m26.34s & -38d18m08.85s & 241.28118 & -42.70393 & 10.96 \\ \addlinespace
    4874147282294634496 & 04h46m26.38s & -33d09m18.77s & 234.89973 & -39.4875 & 10.88 \\ \addlinespace
    4874355845906645120 & 04h44m49.12s & -32d52m40.74s & 234.47078 & -39.77595 & 9.82 \\ \addlinespace
    5297612567846523648 & 08h55m16.79s & -63d11m48.85s & 279.68487 & -11.56833 & 10.59 \\ \addlinespace
    5303348140199025920 & 08h53m18.62s & -59d02m02.45s & 276.24095 & -9.13719 & 9.18 \\ \addlinespace
    5489531880096156416 & 07h53m21.37s & -52d39m13.01s & 266.12717 & -12.61531 & 9.46 \\ \addlinespace
    5490827169212941056 & 07h16m15.58s & -54d16m38.91s & 265.24364 & -18.30189 & 10.37 \\ \addlinespace
    5494465006512656000 & 06h01m47.64s & -59d51m51.34s & 268.69547 & -29.3658 & 10.18 \\ \addlinespace
    5498528840145966464 & 06h55m28.03s & -52d47m36.15s & 262.66243 & -20.71451 & 9.96 \\ \addlinespace
    5510893810476230144 & 07h34m18.14s & -45d16m36.11s & 257.95676 & -11.97777 & 10.54 \\ \addlinespace
    5534999297246168320 & 07h35m22.44s & -44d25m02.43s & 257.26085 & -11.42081 & 10.75 \\ \addlinespace
    5537359398962337280 & 07h56m34.82s & -40d20m27.82s & 255.56685 & -6.02412 & 10.96 \\ \addlinespace
    5539856596020917376 & 08h15m04.77s & -39d58m08.40s & 257.14142 & -2.83572 & 10.98 \\ \addlinespace
    5545190739243196672 & 07h58m07.44s & -35d54m56.94s & 251.92065 & -3.47982 & 10.42 \\ \addlinespace
    5549536180990319616 & 06h25m44.65s & -50d08m47.30s & 258.54324 & -24.4474 & 10.88 \\ \addlinespace
    5551362018831013376 & 06h43m15.47s & -48d39m59.03s & 257.82336 & -21.30686 & 10.56 \\ \addlinespace
    5551565291043498496 & 06h41m26.76s & -48d13m11.51s & 257.2697 & -21.46142 & 10.52 \\ \addlinespace
    5555201689530777344 & 06h28m22.83s & -47d46m42.24s & 256.15226 & -23.43749 & 10.97 \\ \addlinespace
    5556830959605647360 & 06h29m33.11s & -44d15m03.85s & 252.53326 & -22.27004 & 10.21 \\ \addlinespace
    5557022343348187392 & 06h42m10.34s & -44d02m26.04s & 253.06044 & -20.05018 & 10.88 \\ \addlinespace
    5578884070483294976 & 06h52m22.61s & -36d24m18.71s & 246.28301 & -15.52673 & 10.8 \\ \addlinespace
    5584821364554787456 & 07h25m11.33s & -40d23m42.41s & 252.73421 & -11.30575 & 10.32 \\ \addlinespace
    5586241315104190848 & 07h28m03.43s & -38d00m45.45s & 250.81879 & -9.74411 & 9.68 \\ \addlinespace
    5616551552155482880 & 07h18m43.25s & -24d39m26.88s & 237.91749 & -5.40371 & 8.94 \\ \addlinespace
    5618295476367781504 & 07h30m41.52s & -23d59m08.82s & 238.60724 & -2.68067 & 10.91 \\ \addlinespace
    \end{tabularx}
    \label{tab:LOPS2_special_targets}
    \tablefoot{These stars are classified as halo stars based on the kinematic characterisation, and fall in the high signal-to-noise ratio P1 PLATO sample. The P1 sample contains all targets with the additional requirements that $V \leq 11$ and that the random noise level is below 50ppm in one hour.}
\end{table*}

\begin{table*}
    \centering
    \caption{Same as Table \ref{tab:LOPS2_special_targets}, but for the LOPN1 field.}
    \begin{tabularx}{\textwidth}{r|RR|RR|R}
    \textit{Gaia} DR3 ID & RA (hms) & Dec (dms) & Gal. longitude (deg) & Gal. latitude (deg) & V Magnitude\\ 
    \addlinespace
    \hline\hline
    \addlinespace
    1340991529725341056 & 17h11m40.84s & 37d49m55.08s & 61.75032 & 35.17462 & 10.59 \\ \addlinespace
    1342299192648782592 & 17h39m36.23s & 37d10m48.77s & 62.26064 & 29.61662 & 8.36 \\ \addlinespace
    1423516852416948224 & 16h14m57.15s & 49d46m03.04s & 77.50137 & 45.30106 & 10.98 \\ \addlinespace
    1622478459328957696 & 15h57m27.57s & 56d40m02.77s & 88.01797 & 45.85473 & 10.68 \\ \addlinespace
    1644643411153918336 & 15h39m11.90s & 66d48m13.65s & 101.89417 & 42.83402 & 10.31 \\ \addlinespace
    2026374267595492096 & 19h28m53.81s & 28d22m21.21s & 62.28341 & 5.12122 & 10.29 \\ \addlinespace
    2039347061671874944 & 19h10m44.00s & 30d05m46.72s & 62.01238 & 9.43357 & 10.76 \\ \addlinespace
    2051426296414984960 & 19h31m09.21s & 36d09m01.42s & 69.4295 & 8.34166 & 10.2 \\ \addlinespace
    2077092436860985728 & 19h42m06.28s & 41d41m23.01s & 75.38792 & 9.06408 & 10.62 \\ \addlinespace
    2083249324019906048 & 20h20m38.47s & 46d26m29.55s & 83.09395 & 5.60762 & 10.81 \\ \addlinespace
    2104987557947509888 & 18h54m17.01s & 42d59m04.32s & 72.80069 & 17.62626 & 9.93 \\ \addlinespace
    2107126146721252864 & 18h59m16.93s & 45d06m31.93s & 75.21886 & 17.55053 & 10.89 \\ \addlinespace
    2126182469941128960 & 19h26m03.33s & 44d21m35.14s & 76.45543 & 12.88873 & 10.27 \\ \addlinespace
    2142082129629510272 & 19h32m25.97s & 56d36m25.13s & 88.4213 & 17.07098 & 10.32 \\ \addlinespace
    2203746967971153024 & 21h39m15.35s & 60d17m05.38s & 101.18362 & 5.802 & 10.32 \\ \addlinespace
    \end{tabularx}
    \label{tab:LOPN1_special_targets}
\end{table*}
\end{appendix}

\end{document}